\RequirePackage{fix-cm}
\documentclass{ieeetj}[tikz,border=10pt]
\usepackage{cite}
\usepackage{amsmath,amssymb,amsfonts}
\usepackage{algorithmic}
\usepackage{graphicx,color}
\usepackage{textcomp}
\usepackage{xcolor}
\usepackage{hyperref}
\hypersetup{hidelinks}

\usepackage{booktabs}
\usepackage{graphicx}
\usepackage{multirow}
\usepackage{amsmath} 
\usepackage{xcolor}  
\usepackage{paralist}

\usepackage{algorithm,algorithmic}
\usepackage[nowarn,acronyms,nonumberlist,nopostdot,nomain,nogroupskip]{glossaries}
\newacronym{3gpp}{3GPP}{3rd Generation Partnership Project}
\newacronym{itu}{ITU}{International Telecommunication Union}
\newacronym{aerpaw}{AERPAW}{Aerial Experimentation and Research Platform for Advanced Wireless}
\newacronym{wifi}{WiFi}{Wireless Fidelity}
\newacronym{sls}{SLS}{System Level Simulations}
\newacronym{kpis}{KPIs}{Key Performance Indicators}
\newacronym{kpm}{KPM}{key performance metric}
\newacronym{lte}{LTE}{Long Term Evolution}
\newacronym{mpquic}{MPQUIC}{Multi-Path QUIC}
\newacronym{ap}{AP}{access point}
\newacronym{aps}{APs}{access points}
\newacronym{rssi}{RSSI}{received signal strength index}
\newacronym{bs}{BS}{base station}
\newacronym{vho}{VHO}{vertical handover}
\newacronym{ml}{ML}{machine learning}
\newacronym{lstm}{LSTM}{long short term memory}
\newacronym{qos}{QoS}{quality of service}
\newacronym{uma}{UMa}{urban macro}
\newacronym{inh}{InH}{indoor hall}
\newacronym{i2i}{I2I}{Indoor to Indoor}
\newacronym{i2o}{I2O}{Indoor to Outdoor}
\newacronym{o2i}{O2I}{Outdoor to Indoor}
\newacronym{o2o}{O2O}{Outdoor to Outdoor}
\newacronym{los}{LoS}{line-of-sight}
\newacronym{nlos}{NLOS}{non line of sight}
\newacronym{rsrp}{RSRP}{reference signal received power}
\newacronym{rsrq}{RSRQ}{reference signal received quality}
\newacronym{sinr}{SINR}{signal-to-interference-plus-noise Ratio}
\newacronym{pci}{PCI}{physical cell ID}
\newacronym{ta}{TA}{timing advance}
\newacronym{bssid}{BSSID}{basic service set identifier}
\newacronym{ssid}{SSID}{service set identifier}
\newacronym{rl}{RL}{reinforcement learning}
\newacronym{dqn}{DQN}{deep Q-network}
\newacronym{gsm}{GSM}{Global System for Mobile}
\newacronym{nr}{NR}{5G New Radio}
\newacronym{hetnets}{HetNets}{Heterogeneous Networks}
\newacronym{qoe}{QOE}{quality of experience}
\newacronym{gps}{GPS}{Global Positioning System}
\newacronym{swcp}{SWCP}{sticky WiFi client problem}
\newacronym{iot}{IOT}{Internet of Things}
\newacronym{uav}{UAV}{unmanned aerial vehicle}
\newacronym{oran}{ORAN}{Open Radio Access Networks}
\newacronym{rem}{REM}{radio environment map}
\newacronym{rma}{RMa}{Rural Macro Environment}
\newacronym{lw}{LW}{Lake Wheeler}
\newacronym{rf}{RF}{Radio Frequency}
\newacronym{rdz}{RDZ}{radio dynamic zone}
\newacronym{son}{SON}{Self-Organizing Networks}
\newacronym{nlp}{NLP}{natural language processing}
\newacronym{fspl}{FSPL}{free space path loss}
\newacronym{lwsrd}{LWSRD}{linear warmup and square-root decay}
\newacronym{gat}{GAT}{graph attention network}
\newacronym{gan}{GAN}{generative adversarial network}
\newacronym{mae}{MAE}{mean absolute error}
\newacronym{rmse}{RMSE}{root mean square error}
\newacronym{gpr}{GPR}{Gaussian process regression}
\newacronym{mse}{MSE}{mean square error}
\newacronym{cqi}{CQI}{channel quality indicator}
\newacronym{nsa}{NSA}{non-standalone}
\newacronym{ss}{SS}{synchronization signal}
\newacronym{ssm}{SSM}{state-space model}
\newacronym{cnn}{CNN}{convolutional neural network}
\newacronym{gru}{GRU}{gated recurrent unit}
\newacronym{ssim}{SSIM}{structural similarity index metric}
\newacronym{fsim}{FSIM}{feature similarity index metric}
\newacronym{mcs}{MCS}{modulation and coding scheme}
\newacronym{ri}{RI}{rank indicator}
\newacronym{urllc}{URLLC}{ultra reliable low latency communication}
\newacronym{wb-cqi}{WB-CQI}{wide band-CQI}
\newacronym{dnn}{DNN}{deep neural network}
\newacronym{vit}{ViT}{vision transformer}
\newacronym{pdsch}{PDSCH}{physical downlink shared channel}
\newacronym{gpu}{GPU}{graphics processing unit }
\usepackage{xurl}

\renewcommand{\baselinestretch}{0.96}
\def\BibTeX{{\rm B\kern-.05em{\sc i\kern-.025em b}\kern-.08em
    T\kern-.1667em\lower.7ex\hbox{E}\kern-.125emX}}
\AtBeginDocument{\definecolor{tmlcncolor}{cmyk}{0.93,0.59,0.15,0.02}\definecolor{NavyBlue}{RGB}{0,86,125}}

\usepackage{tikz}
\usetikzlibrary{arrows.meta, angles, quotes, shapes.geometric}

\def\OJlogo{\vspace{-6pt}\hbox{$<$Submitted for possible publication in IEEE JSTEAP.$>$}}
\def\seclogo{\vspace{10pt}$<$Submitted for possible publication in IEEE JSTEAP.$>$}

\def\authorrefmark#1{\ensuremath{^{\textbf{#1}}}}

\begin{document}
\receiveddate{XX Month, XXXX}
\reviseddate{XX Month, XXXX}
\accepteddate{XX Month, XXXX}
\publisheddate{XX Month, XXXX}
\currentdate{XX Month, XXXX}
\doiinfo{}

\markboth{}{}

\title{AI-Enabled Wireless Propagation Modeling and Radio Environment Maps for 5G Aerial Wireless Networks}

\author{Gautham Reddy\authorrefmark{1}, K\"ur\c{s}at Tekb\i y\i k\authorrefmark{2}, Bryton Petersen\authorrefmark{3},\\ Antoine Lesage-Landry\authorrefmark{2}, Gunes Karabulut Kurt\authorrefmark{2}, and  Ismail G\"{u}ven\c{c}\authorrefmark{1}}
\affil{Department of Electrical and Computer Engineering, NC State University, Raleigh, NC 27606, USA}
\affil{Department of Electrical Engineering, Polytechnique Montréal and Poly-Grames Research Centre,  Montréal, QC H3T 0A3, Canada}
\affil{Idaho National Laboratory, Idaho Falls, ID 83415, USA}
\corresp{Corresponding author: Gautham Reddy (email: greddy2@ncsu.edu).}
\authornote{This work was supported in part by the INL Laboratory Directed Research Development (LDRD) Program under BMC No. 264247, Release No. 26 on BEA's Prime Contract No. DE-AC07-05ID14517, the NSF award CNS-2332835, and the Natural Sciences and Engineering Research Council (NSERC) of Canada Alliance grant ALLRP 579869-22 .}

\begin{abstract}  
With the gaining prominence of aerial mobility applications, their success depends on the seamless integration of terrestrial and non-terrestrial network connectivity.
However, providing reliable connectivity from terrestrial telecommunication networks remains challenging due to multi-cell interference from \glspl{bs} under \gls{los} conditions to \glspl{uav}, coverage holes caused by antenna sidelobe degradation, localized multipath fading effects, and the high-speed dynamics of aerial users.
To model such complexities, often exacerbated by sparse real-world data, this work proposes a dual-stage \gls{rem} framework.
Our approach physically decouples the channel modeling, where a spatial Transformer first anchors the deterministic, large-scale path loss geometry, while a \gls{gru} subsequently extrapolates the stochastic, localized fast-fading deviations. 
By reformulating 3D spatial interpolation as a 1D radial sequence prediction task, the framework inherently aligns with the physics of propagation.
We evaluate the proposed framework against state-of-the-art baselines, including 3D Kriging, UNet, Mamba, and Inception, using empirical 5G datasets.
The results demonstrate improved intra-site generalization across diverse altitudes, user dynamics, and \gls{rsrp} datasets, achieving signal-strength predictions with errors near $3$~dB and \gls{rem} spatial-similarity indices exceeding $0.75$.  
Finally, we examine the influence of \glspl{rem} and channel rank conditions on \gls{uav} channel quality, underscoring the necessity of reliable channel modeling for robust aerial connectivity.

\end{abstract}



\maketitle

\glsresetall

\section{INTRODUCTION}
\IEEEPARstart{T}{he} rapid proliferation of \glspl{uav} is fundamentally reshaping wireless connectivity by extending the traditional propagation space in the vertical dimension. This integration of \glspl{uav} into future wireless ecosystems also presents new challenges for network planning. Specifically, the demand for reliable three-dimensional (3D) connectivity must often be met using infrastructure designed primarily for terrestrial users. Unlike ground users, aerial nodes operate in a dynamic 3D environment characterized by high-probability \gls{los} links and interference from terrestrial \gls{bs} sidelobes. These unique conditions necessitate new frameworks for interference mitigation, reliable spectrum management, and power optimization to ensure robust \gls{uav} connectivity.

\begin{figure*}
    \includegraphics[width=\linewidth]{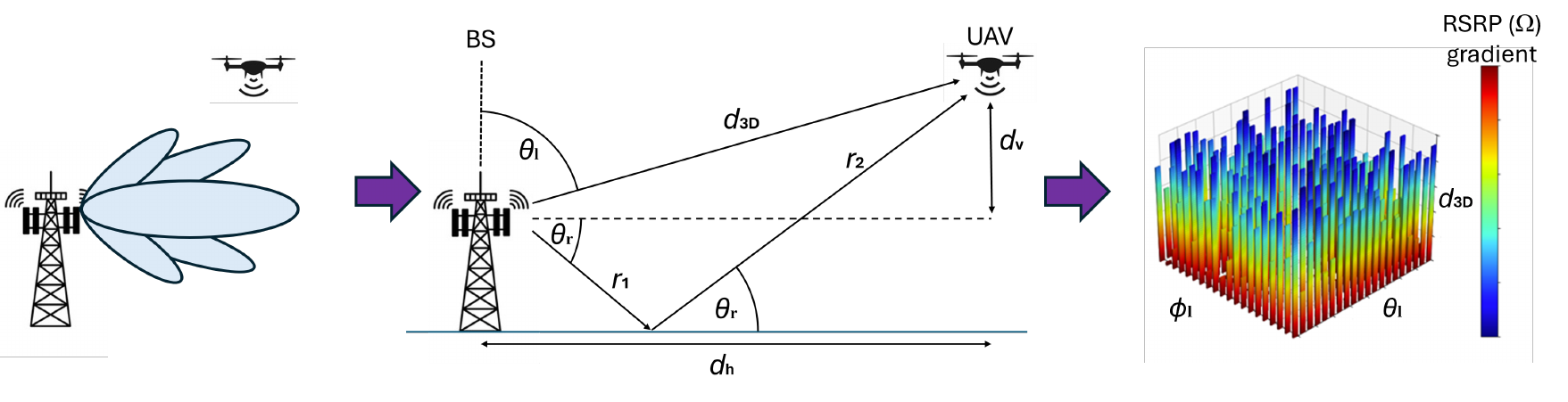}
	\centering
    \caption{The ground-to-UAV propagation environment is modeled using a two-ray channel approach, with received signals visualized as radial RSRP gradient sequences originating from the \gls{bs} for every 3D elevation ($\theta_{\rm l}$) and azimuth ($\phi_{\rm l}$) angle.}
    \label{fig:two_ray_model}
\end{figure*}

To address these challenges, \glspl{rem} have emerged as critical enablers for proactive resource allocation, interference management, and spatial spectrum utilization. \glspl{rem} capture the complex interactions of electromagnetic propagation in a spatial map and, therefore, make wireless network management proactive rather than reactive through awareness of signal quality metrics such as \gls{rsrp} or \gls{sinr} across a volumetric 3D coverage space.

Utilizing high-fidelity 3D \glspl{rem} in aerial communication systems can provide numerous benefits. First, \glspl{rem} enable predictive signal strength modeling along flight paths, allowing for proactive handovers and improved network coverage. 
Second, they facilitate dynamic spectrum access, optimizing utilization by identifying non-allocated frequency bands within specific spatial regions, forming \glspl{rdz}.
Finally, flight paths can be generated through the dual optimization of propulsion and communication, ensuring the \gls{uav} navigates energy-efficient routes that satisfy reliable \gls{qos} link constraints. 

\subsection{Related Works}

Previously, \gls{rem} construction methods have relied on either deterministic propagation models or statistical interpolation techniques~\cite{feng2025recent}. 
Deterministic methods, such as ray tracing~\cite{yun2015ray}, offer physics-based predictions but require comprehensive 3D environmental data and substantial computational resources. In the absence of such detailed inputs, these methods often fail to capture site-specific shadowing despite their high computational overhead.
Statistical approaches, such as the Kriging method~\cite{SungJoonKriging}, provide a mathematical basis for interpolation, but they are poorly scalable to large datasets because they often exhibit cubic computational complexity~\cite{van2020cluster}.

To overcome these scalability and accuracy limitations, data-driven \gls{dnn} methods are widely investigated. For example, RadioUNet~\cite{levie2021radiounet} is adapted to treat radio mapping as an image-to-image translation task and utilizes UNet encoder-decoder architectures to capture global spatial features and create \glspl{rem} from propagation medium images. Recently, in~\cite{hehn2023transformer}, a \gls{vit}-based method has been proposed for a similar image-to-map task. It achieves better performance than \glspl{cnn} by leveraging attention mechanisms to model correlations among samples. 
Despite their performance, a significant portion of prior literature treats \gls{rem} construction as a computer vision image inpainting problem, utilizing \glspl{cnn}~\cite{CNN_3DREM}, \glspl{gan}~\cite{GAN_3DREM} and diffusion models~\cite{RadioLAM} to predict signal values on a discretized 2D/3D grid. However, estimation from a propagation medium image is challenging and unrealistic for aerial networks operating in a dynamic 3D environment, with limited \gls{los} obstructions, narrow \gls{bs} antenna sidelobes, and rapid \gls{uav} mobility. To close this gap, we propose a paradigm shift from a grid-based image view to a continuous radial sequence view, seen in Fig.~\ref{fig:two_ray_model}. Rather than estimating pixels on a map, our framework formulates the propagation path as a 1D spatial sequence extending radially from the base station to the \gls{uav}.

Therefore, we propose a physics-inspired framework that explicitly captures the effects of the transmitter, channel, and receiver parameters by formulating \gls{rem} generation as a sequence-to-sequence mapping problem. The proposed approach is better suited to dynamic environments, such as aerial networks, because it primarily relies on spatial measurements and the \gls{uav}'s position.
Before discussing the details of our method, we briefly explain other sequence-to-sequence models that are candidate solutions for the same problem. For example, recently proposed Mamba architectures use \glspl{ssm} to achieve linear training complexity, and a global memory as an efficient alternative to the quadratic scaling of transformers~\cite{gu2024mamba}. Similarly, Inception networks~\cite{szegedy2015going} use parallel convolutions with varying kernel sizes to capture temporal features at different scales. Additionally, the physics-inspired Kolmogorov-Arnold network (PIKAN)~\cite{tekbiyik2025pikan} provides symbolic expressions of the propagation medium, making it interpretable and explainable. Advancements of such sequence models, including our prior work TransfoREM~\cite{TransfoREM_ICC2026}, have demonstrated the theoretical and computational advantages of grid-less, 1D radial sequence modeling over traditional 2D patch-based spatial grids for the aerial \gls{rem} problem. 

More recent literature has explored structural interpolation methods for 3D \glspl{rem}, such as evaluating Kriging against matrix completion \cite{Rahman_DySPAN2024}. By partitioning the 3D space into discrete grids, these matrix-based approaches effectively reconstruct maps by leveraging the low-rank properties of the environment. However, while highly effective for structurally smooth and stationary spaces, these grid-dependent operations predominantly assume a static radio map with fixed small-scale fading behavior. This assumption inherently masks the highly dynamic, user-specific \gls{rsrp} variation trends inherent to aerial mobility applications~\cite{rahmanUAV3dspectrum}. Furthermore, conventional spatio-temporal hybrid models typically rely on feature concatenation to track map-level temporal changes, a computationally demanding approach. To address these limitations and move beyond static grid formulations, our proposed framework introduces a paradigm of physical decoupling. By viewing the 3D \gls{rem} in a continuous spherical coordinate perspective, we avoid the constraints of discrete matrix formulations. We then utilize a spatial Transformer to evaluate macroscopic channel effects from the \gls{bs} perspective and a \gls{gru} estimator to act as a stochastic extrapolator for unobservable micro-environmental fast-fading. We thus establish an architecture in which various learning implementations can efficiently map the decomposed components. In this manner, we provide a unified architecture capable of generating \glspl{rem} from sparse, temporally varying measurements and extending it for user-specific \gls{rsrp} prediction in continuous 3D aerial networks.

\subsection{Contributions}
The specific contributions of this article are:
\begin{enumerate}[{C}1.]
    \item \textbf{Cascaded physics-based learning:} We propose a theoretically motivated, decoupled \gls{rem} architecture that mirrors the physical superposition of wireless channels. By formulating large-scale spatial propagation as an attention-driven sequence and fast-fading multipath as a temporally extrapolated stochastic process, we overcome the limitations of standard spatial-temporal radio environment predictive models.
    \item \textbf{Spatial and temporal correlation analysis:} 
    Through a detailed analysis of 3D signal strength data, we demonstrate that aerial signal propagation exhibits strong 1D radial spatial correlation and sequential temporal fast-fading variation. These inherent physical characteristics directly motivate our proposed \gls{rem} architecture.
    \item \textbf{Intra-site generalization across different datasets:} We evaluate our model on extensive real-world 5G datasets collected at the AERPAW platform~\cite{Dataset18, Dataset23, Dataset24}, using PawPrints, Nemo, and Quectel devices within the same environment. We show that our method outperforms related benchmarks, including 3D Kriging, UNet, Mamba, and Inception, especially in \gls{rem} reconstruction and \gls{rsrp} prediction from sparse measurements across varying altitudes and \gls{uav} dynamics.
\end{enumerate}

\subsection{Outline}
The rest of the paper is organized as follows. 
Section~\ref{sec:SystemModel} details the radio propagation model, real-world data collection, and analysis. 
Section~\ref{sec:ModelArchitecture} then discusses the Transformer and \gls{gru} models used to generate \glspl{rem}. 
In Section~\ref{sec:Results}, we present the experimental results and include a case study on \gls{uav} channel quality prediction using \glspl{rem} in Section~\ref{sec:Results}-\ref{sec:CaseStudy}.
Finally, Section~\ref{sec:Conclusion} concludes this work.


\section{RADIO PROPAGATION MODEL}
\label{sec:SystemModel}

Past research in wireless propagation includes deterministic models tailored to specific spatial contexts, ranging from satellite-to-ground links to urban macro-cells and near-field communications. This work focuses on the ground-to-air propagation channel, which governs \gls{uav} communication scenarios. In contrast to the dense multi-path environments of urban centers, rural scenarios prominently exhibit the two-ray path loss model, which characterizes the coupling of a direct \gls{los} ray and a ground-reflected component. Building upon the framework established in~\cite{SungJoonKriging}, the propagation geometry between the \gls{bs} and the \gls{uav} is illustrated in Fig.~\ref{fig:two_ray_model}. Within this geometry, the two-ray path loss is decomposed into a deterministic \gls{los} component and a stochastic fading factor.

Consider that the \gls{bs}, \gls{uav} locations $\mathbf{l}^{\mathrm{bs}}$, $\mathbf{l}^{\mathrm{uav}}$ are given by $\{\psi^{\mathrm{bs}}, \omega^{\mathrm{bs}}, h^{\mathrm{bs}}\}$ and $\{\psi^{\mathrm{uav}}, \omega^{\mathrm{uav}}, h^{\mathrm{uav}}\}$, respectively, with $\psi, \omega,$ and $h$ denoting the latitude, longitude, and altitude of the locations. Then, the relative separation distance and angular orientations are calculated as follows:
\begin{equation}
    \label{eq:3Dgeometery}
    \begin{aligned}
    d_{\mathrm{h}}\left(\mathbf{l}^{\mathrm{bs}}, \mathbf{l}^{\mathrm{uav}}\right) &= A \times \arccos \left(\sin \psi^{\mathrm{uav}} \sin \psi^{\mathrm{bs}}\right. \\
    &\quad \left.+\cos \psi^{\mathrm{uav}} \cos \psi^{\mathrm{bs}} \cos \left(\omega^{\mathrm{bs}}-\omega^{\mathrm{uav}}\right)\right), \\
    d_{\mathrm{v}}\left(\mathbf{l}^{\mathrm{bs}}, \mathbf{l}^{\mathrm{uav}}\right) &=\left|h^{\mathrm{bs}}-h^{\mathrm{uav}}\right|, \\
    d_{3 \rm D}\left(\mathbf{l}^{\mathrm{bs}}, \mathbf{l}^{\mathrm{uav}}\right) &=
    \sqrt{d_{\mathrm{h}}\left(\mathbf{l}^{\mathrm{bs}}, \mathbf{l}^{\mathrm{uav}}\right)^2 + d_{\mathrm{v}}\left(\mathbf{l}^{\mathrm{bs}}, \mathbf{l}^{\mathrm{uav}}\right)^2}, \\
    \theta_{\rm l} &= \tan^{-1}\left( d_{\mathrm{h}}/d_{\mathrm{v}}\right), \\
    \theta_{\rm r} &= \tan^{-1}\left(\left(h^{\mathrm{bs}}+h^{\mathrm{uav}}\right)/d_{\mathrm{h}}\right), \\
    \end{aligned}
\end{equation}
where $A$ is the radius of the Earth ($\approx$6,378,137 m), $d_{\mathrm{h}}$ is the horizontal separation distance, $d_{\mathrm{v}}$ is the vertical separation distance, $d_{3 \rm D}$ is the \gls{los} separation distance, $\theta_{\rm l}$ is the \gls{los} elevation angle and $\theta_{\rm r}$ is the angle of reflection with respect to the ground.

The path loss for the two-ray model is expressed as:
\begin{align}
    & PL_{\rm trm}\left(\mathbf{l}^{\mathrm{bs}}, \mathbf{l}^{\mathrm{uav}}\right)
    =\left(\frac{c}{4 \pi f_{\rm c}}\right)^2 \left\lvert\, \underbrace{\frac{\sqrt{\mathrm{G}_{\mathrm{bs}}\left(\phi_{\rm l}, \theta_{\rm l}\right) \mathrm{G}_{\mathrm{uav}}\left(\phi_{\rm l}, \theta_{\rm l}\right)}}{d_{3\mathrm{D}}}}_{\text{LoS signal }}\right. \nonumber \\
    &\quad\quad\quad +\left.\underbrace{\frac{\Gamma\left(\theta_{\rm r}\right) \sqrt{\mathrm{G}_{\mathrm{bs}}\left(\phi_{\rm r}, \theta_{\rm r}\right) \mathrm{G}_{\text {uav}}\left(\phi_{\rm r}, \theta_{\rm r}\right)} {\rm e}^{-j \Delta \tau}}{r_1+r_2}}_{\text {ground reflected signal }}\right|^2,
    \label{eq:2ray_loss}
\end{align}
where $\mathrm{G}_{\mathrm{bs}}\left(\phi, \theta\right)$, and $\mathrm{G}_{\mathrm{uav}}\left(\phi, \theta\right)$ are the \gls{bs} and \gls{uav} antenna gains as functions of elevation and azimuth angles, $c$ is the speed of light, $f_{\rm c}$ is the center frequency, $\Gamma\left(\theta_{\rm r}\right)$ is the ground reflection co-efficient, $\Delta \tau = \left(\left( 2 \pi \left(r_1 + r_2 - d_{\rm 3D}\right)\right) c/ f_{\rm c}\right)$ is the phase difference between the two paths, and $r_{\mathrm{1}} = h^{\mathrm{bs}}/\sin{\theta_{\rm r}}$, $r_{\mathrm{2}} = h^{\mathrm{uav}}/\sin{\theta_{\rm r}}$ are the lengths of the reflected rays.

\begin{figure*}
    \centering
    \includegraphics[width=0.85\linewidth]{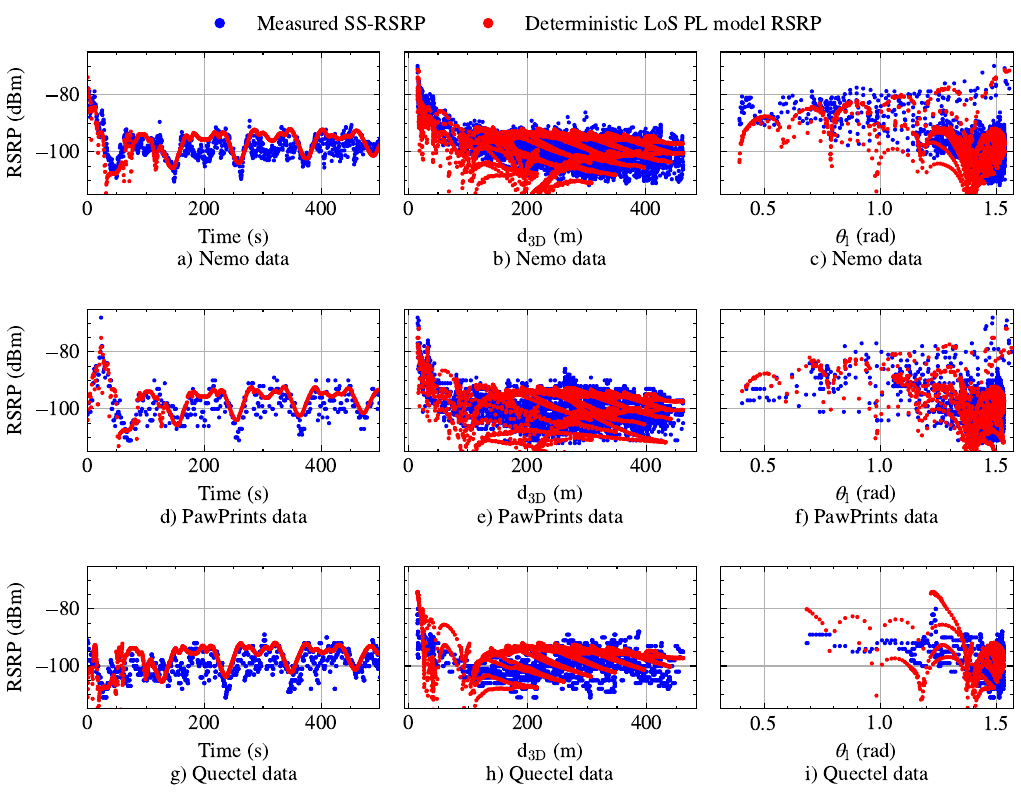}
    \caption{\gls{rsrp} variation from the measured and deterministic \gls{los} path loss model across different channel parameters. (The Quectel dataset, limited to a 30~m altitude slice, has a smaller spread in the elevation angle domain.)}
    \label{fig:datasets_plot}
\end{figure*}

By normalizing the expression with respect to the \gls{los} component, (\ref{eq:2ray_loss}) can be expressed as a product of Friis \gls{fspl} equation and a multipath coupling factor:
\begin{equation}
    \begin{split}
    & PL_{\rm trm}\left(\mathbf{l}^{\mathrm{bs}}, \mathbf{l}^{\mathrm{uav}}\right) = \underbrace{\left(\frac{c}{4 \pi f_{\rm c} d_{\rm 3D}}\right)^2 G_{\rm bs}\left(\phi_{\rm l}, \theta_{\rm l}\right) G_{\rm uav}\left(\phi_{\rm l}, \theta_{\rm l}\right)}_{\text {Friis Equation }} \\
    & \times \underbrace{\left|1+\frac{\Gamma\left(\theta_{\rm r}\right) \sqrt{G_{\rm bs}\left(\phi_{\rm r}, \theta_{\rm r}\right) G_{\rm uav}\left(\phi_{\rm r}, \theta_{\rm r}\right)} {\rm e}^{-j \Delta \tau} \cdot d_{3 \rm D}}{\left(r_1+r_2\right) \sqrt{G_{\rm bs}\left(\phi_{\rm l}, \theta_{\rm l}\right) G_{\rm uav}\left(\phi_{\rm l}, \theta_{\rm l}\right)}}\right|^2}_{\text {Multipath Coupling Factor }}.
    \label{eq:2ray_loss_split}
    \raisetag{60pt}
    \end{split}
\end{equation}
The fundamental parameters governing path loss behavior are captured in (\ref{eq:2ray_loss_split}), and it serves as the theoretical foundation for our \gls{rem} construction.
When expressed on the dB scale,~(\ref{eq:2ray_loss_split}) can further be split as:
\begin{equation}
    \label{eq:2ray_loss_split_dB}
    \begin{split}
        &PL_{\rm trm}^{\rm dB}\left(\mathbf{l}^{\mathrm{bs}}, \mathbf{l}^{\mathrm{uav}}\right) =
        \underbrace{20\log_{10}\left(c/4 \pi\right) - 20\log_{10}\left(f_{\rm c}\right)}_{\text {LoS path loss}} \\
        &\qquad\qquad\quad\underbrace{- 20\log_{10}\left(d_{\rm 3D}\right) + 10\log_{10}\left(G_{\rm bs}\left(\phi_{\rm l}, \theta_{\rm l}\right)\right)}_{\text {LoS path loss}} \\
        &\qquad\qquad\quad\underbrace{+ 10\log_{10}\left(G_{\rm uav}\left(\phi_{\rm l}, \theta_{\rm l}\right)\right) + 20\log_{10}}_{\text {Fast-fading factor} }\\
        &  \underbrace{\left(\left|1+\frac{\Gamma\left(\theta_{\rm r}\right) \sqrt{G_{\rm bs}\left(\phi_{\rm r}, \theta_{\rm r}\right) G_{\rm uav}\left(\phi_{\rm r}, \theta_{\rm r}\right)} {\rm e}^{-j \Delta \tau} \cdot d_{3 \rm D}}{\left(r_1+r_2\right) \sqrt{G_{\rm bs}\left(\phi_{\rm l}, \theta_{\rm l}\right) G_{\rm uav}\left(\phi_{\rm l}, \theta_{\rm l}\right)}}\right| \right)}_{\text {Fast-fading factor}},
        \raisetag{80pt}
    \end{split}
\end{equation}
with a large scale \gls{los} component and a fast-fading factor dependent on the receiver gain and multipath coupling.
The fast-fading factor can also account for additional multipath components that materialize over different regions in 3D space. 
This stagewise path loss behavior can be equivalently implemented as a multi-stage \gls{ml} training strategy. 
With the initial stages focused on modeling the spatial large-scale fading as a function of the \gls{uav}'s position, and the subsequent stage targeting the estimation of the fast-fading factor. 

\begin{figure}
    \centering
    \includegraphics[width=0.9\linewidth]{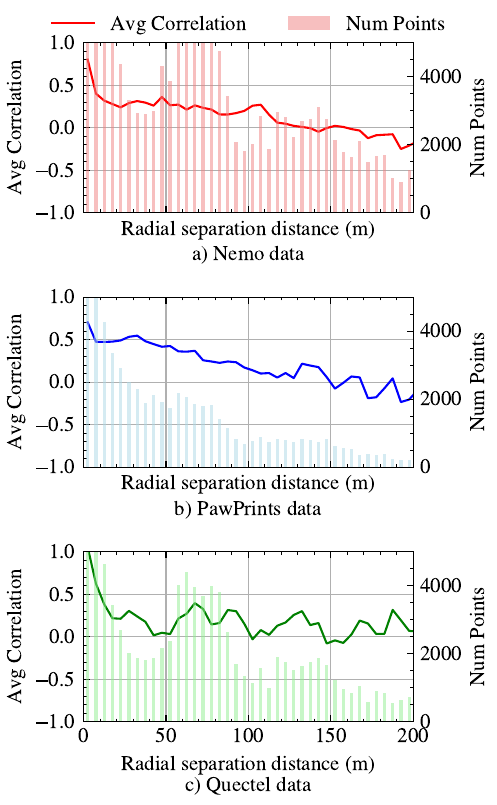}
    \caption{The spatial correlation of measured \glspl{rsrp} values from a) Nemo, b) PawPrints, and c) Quectel datasets in the spherical coordinate system. }
    \label{fig:SpatialCorrelation}
\end{figure}

\subsection{5G Measurement Datasets}\label{SubSec:Datasets}
In this subsection, we utilize the 5G datasets captured from the \gls{aerpaw} Lake Wheeler testbed located in a rural terrain. The outdoor testbed consists of a 5G \gls{nsa} \gls{bs} built with Ericsson equipment, together with open-source programmable \glspl{uav} to conduct aerial wireless experiments. We utilize 5G NR \glspl{kpm} obtained from the datasets in~\cite{Dataset18, Dataset23, Dataset24}, which are collected via multiple \gls{uav}-mounted receivers over the same \gls{uav} trajectory. The measurement campaign employs three distinct hardware configurations: 1) a Samsung S21 smartphone utilizing the PawPrints custom Android application, 2) a Samsung S23+ running the Keysight Nemo Handy software, and 3) a programmable Quectel modem. All devices operated on \gls{nr} carrier within the n77 band ($3.4$~GHz) with a $100$~MHz channel bandwidth. An FCC experimental license (Call sign: WK2XQH) is used to transmit in this band from the \gls{aerpaw} tower. The device \gls{rsrp} reporting intervals were configured at $1.0$~s, $0.5$~s, and $0.25$~s for the PawPrints, Nemo, and Quectel devices, respectively.
Notably, the Quectel dataset is restricted to a $30$~m altitude slice, whereas the PawPrints and Nemo datasets include measurements at both $30$~m and $50$~m around the \gls{bs}. The Quectel dataset distinctly includes two trials along the same 30 m trajectory with fixed \gls{uav} yaw angles of 45° and 315°. We leverage this specific variation to assess the \gls{rem} generality across different fast-fading traces caused by changes in \gls{uav} orientation.

The \gls{ss}-\gls{rsrp} $\Omega_{\mathrm{SS-RSRP}}$ reported by these devices represents the average power of the resource elements carrying the secondary synchronization signals, which we model as:
\begin{equation}
    \Omega_{\mathrm{SS-RSRP}} = P_{\mathrm{TX-SS}}^{\rm dB} + PL_{\rm trm}^{\rm dB} + w,
    \label{eq:rsrp_power}
\end{equation}
where $P_{\mathrm{TX-SS}}^{\rm \left(dB\right)} = 10\log_{10}\left( P_{\rm T}/\left(N_{\rm PRB}N_{\rm SC}\right)\right)$ represents the normalized power per \gls{ss} resource element in the \gls{nr} grid. 
Here, $P_{\rm T}$ denotes the total transmit power, $N_{\rm PRB}$ is the number of physical resource blocks, $N_{\rm SC}$ is the number of subcarriers per resource block, and $w$ accounts for the additional shadowing and noise effects.

To compare \gls{rsrp} variations against the deterministic perspective, we utilize the \gls{los} path loss component from (\ref{eq:2ray_loss_split_dB}), incorporating predetermined \gls{bs} antenna gain within (\ref{eq:rsrp_power}) to calculate the deterministic \gls{rsrp}. Fig.~\ref{fig:datasets_plot} illustrates the \gls{rsrp} fluctuations for each dataset across time, distance $d_{\mathrm{3D}}$, and elevation angle $\theta_{\rm l}$, comparing measured values against the deterministic \gls{los} path loss model values. The empirical data from each device exhibit distinct variation trends, likely stemming from hardware-specific implementations of signal strength calculation, L1/L3 filtering, and measurement averaging.
While the \gls{los} model closely captures large-scale spatial trends, it is inherently limited in its ability to account for the scene-specific shadowing and stochastic small-scale fading observed in the experimental data.

\subsection{Spatial Signal Correlation Properties}
\label{subsec:CorrelationAnalysis}

With the path loss decomposed into a \gls{los} component and a stochastic fast-fading factor in (\ref{eq:2ray_loss_split_dB}), the \gls{los} component exhibits a strong correlation in the spherical domain, driven by the radial nature of signal propagation and the \gls{bs} antenna's spherical gain pattern. In contrast, the fast-fading factor demonstrates a localized correlation with temporally and spatially adjacent samples. This subsection quantifies such correlation properties across the three experimental datasets. 

\subsubsection{RSRP correlation in the spherical coordinate system}

Considering the \gls{rsrp} gradient along radial directions originating from the \gls{bs} as seen in Fig.~\ref{fig:two_ray_model}, we calculate the spatial correlation using the following procedure:
\begin{itemize}
    \item \textbf{Angular Binning:} The 3D environment is partitioned into a spherical grid centered at the \gls{bs} with elevation and azimuth bins of size $0.05$~radians. The grid contains $\frac{\pi}{0.05}\times\frac{2\pi}{0.05} = 7895$ angular bins, and \gls{rsrp} measurements within each angular bin are then grouped.
    \item \textbf{Correlation Computation:} Within each angular group, the pairwise \gls{rsrp} correlations are computed as $r[m,n] = \frac{(\Omega_m - \bar{\Omega})(\Omega_n - \bar{\Omega})}{\sigma^2}$, where $\Omega_m, \Omega_n$ are the \gls{rsrp} observations at $m^{\rm th}$ and $n^{\rm th}$ positions respectively, and $\bar\Omega$ and $\sigma^2$ are the mean and covariance of all \gls{rsrp} measurements in the dataset. The pairwise correlation values $r[m,n]$ are then categorized based on the radial separation between the measurement pairs $d_{\rm{3D}}[m,n] = |d_{\rm{3D}}(\mathbf{l}^{\mathrm{uav}}_m, \mathbf{l}^{\mathrm{bs}}) - d_{\rm{3D}}(\mathbf{l}^{\mathrm{uav}}_n, \mathbf{l}^{\mathrm{bs}})|$, into radial separation bins of size $5$~m.
    \item \textbf{Statistical Aggregation:} Finally, the correlation coefficients within each radial separation bin are averaged across all angular bins to generate a composite \gls{rsrp} correlation profile as a function of radial distance. 
\end{itemize}

Fig.~\ref{fig:SpatialCorrelation} shows the average pairwise spatial correlation of \gls{rsrp} along the radial directions originating from the \gls{bs} in a spherical coordinate system. 
Here, the PawPrints data exhibits the most gradual decay in correlation across distance, with values close to $0.5$ up to $50$~m separation.
It benefits from longer averaging intervals between reports to minimize fast-fading trends. 
The Nemo dataset exhibits a quicker dropoff in correlation, partly affected by its smaller averaging window and decimal-level \gls{rsrp} reports, while the Quectel dataset exhibits a significantly narrower correlation range. 
The Quectel data, with the smallest reporting interval, has sufficient radial bin occupancy but is limited to data profile from $30$~m altitude, thus exhibiting more spiky correlation behavior.  
Overall, the three datasets with sparse 3D sampling are well correlated over radial directions in spherical space, which benefits the mapping of large-scale path loss behavior. 

\begin{figure}
    \includegraphics[width=\linewidth]{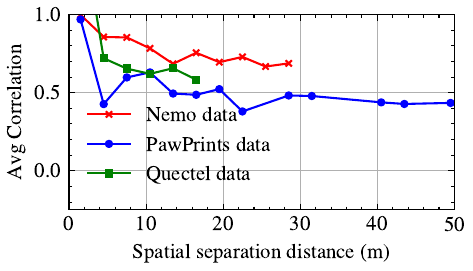}
	\centering
	\caption{Correlation behavior of the current and prior fast-fading factors obtained from a rolling window of five samples aggregated by spatial separation distance.}
    \label{Fig:SeqnCorrelation}
\end{figure}

\subsubsection{Fast-fading factor correlation in spatio-temporal domain}

According to~(\ref{eq:2ray_loss_split_dB}) and~(\ref{eq:rsrp_power}), the fast-fading factor denoted as $\nu$ including regional shadowing $w$, can be obtained by removing the \gls{los} path loss and $P_{\mathrm{TX-SS}}^{\rm{dB}}$ from the received \gls{rsrp}, as $\nu = \Omega - P_{\mathrm{TX-SS}} - PL_{\rm{LoS}}^{\rm{dB}}$. The fast-fading factors thus obtained from adjacent spatial samples are evaluated for correlation and aggregated based on spatial separation distance. Given the maximum \gls{uav} speed of $10$~m/s and differing reporting intervals of up to $1$~s, we evaluate the correlation of the fast-fading factors using a sliding window of the five most recent samples.
Accounting for recent sequential samples is essential due to the coherence distance limitations of fast-fading factors over extended intervals.

The pairwise correlation of fast-fading values relative to position $i$ with the five preceding measurements is calculated as $r[i,m] = \frac{(\nu_i)(\nu_{i-m})}{\sigma^2_\nu}$ for $ m \in \{1,2,..,5\}$, where $\nu_i$ is the $i^{\rm th}$ fast-fading value and $\sigma^2_\nu$ is the covariance of all fast-fading values in the dataset. 
The fast-fading values exhibit a Gaussian distribution with their mean near zero across all three datasets.
These pairwise correlation values are grouped based on spatial separation distance $d_{\rm{3D}}(\mathbf{l}^{\mathrm{uav}}_i, \mathbf{l}^{\mathrm{uav}}_{i-m})$ from~(\ref{eq:3Dgeometery}).
The total set of correlation values across all separation distances is binned with a step size of $3$~m and averaged within each distance bin.

This analysis captures the localized spatial correlation of fast-fading values in varying radii (PawPrints: $5\times1~\rm{s}\times10~\rm{m/s} \simeq 50~\rm{m}$, Nemo: $5\times0.5~\rm{s}\times10~\rm{m/s} \simeq 25~\rm{m}$ and Quectel: $5\times0.25~\rm{s}\times10~\rm{m/s} \simeq 13~\rm{m}$) depending on the reporting intervals.
Fig.~\ref{Fig:SeqnCorrelation} illustrates this behavior with varying degrees of correlation significance.
The Nemo dataset is the most well-correlated among the three, owing to its non-quantized decimal-valued \gls{rsrp} measurements.
Both PawPrints and Quectel provide smoothly filtered and integer-level quantized \gls{rsrp} reports, which fail to capture the fine-grained fast-fading effects.
The Quectel dataset benefits from shorter reporting intervals, which better preserves the correlation nature. 
Improved spatio-temporal fast-fading correlation behavior enables the robust estimation of localized multipath fading and orientation-affected receiver gain changes~(\ref{eq:2ray_loss_split_dB}) over short intervals.

\begin{figure}[b]
	\includegraphics[width=0.9\linewidth]{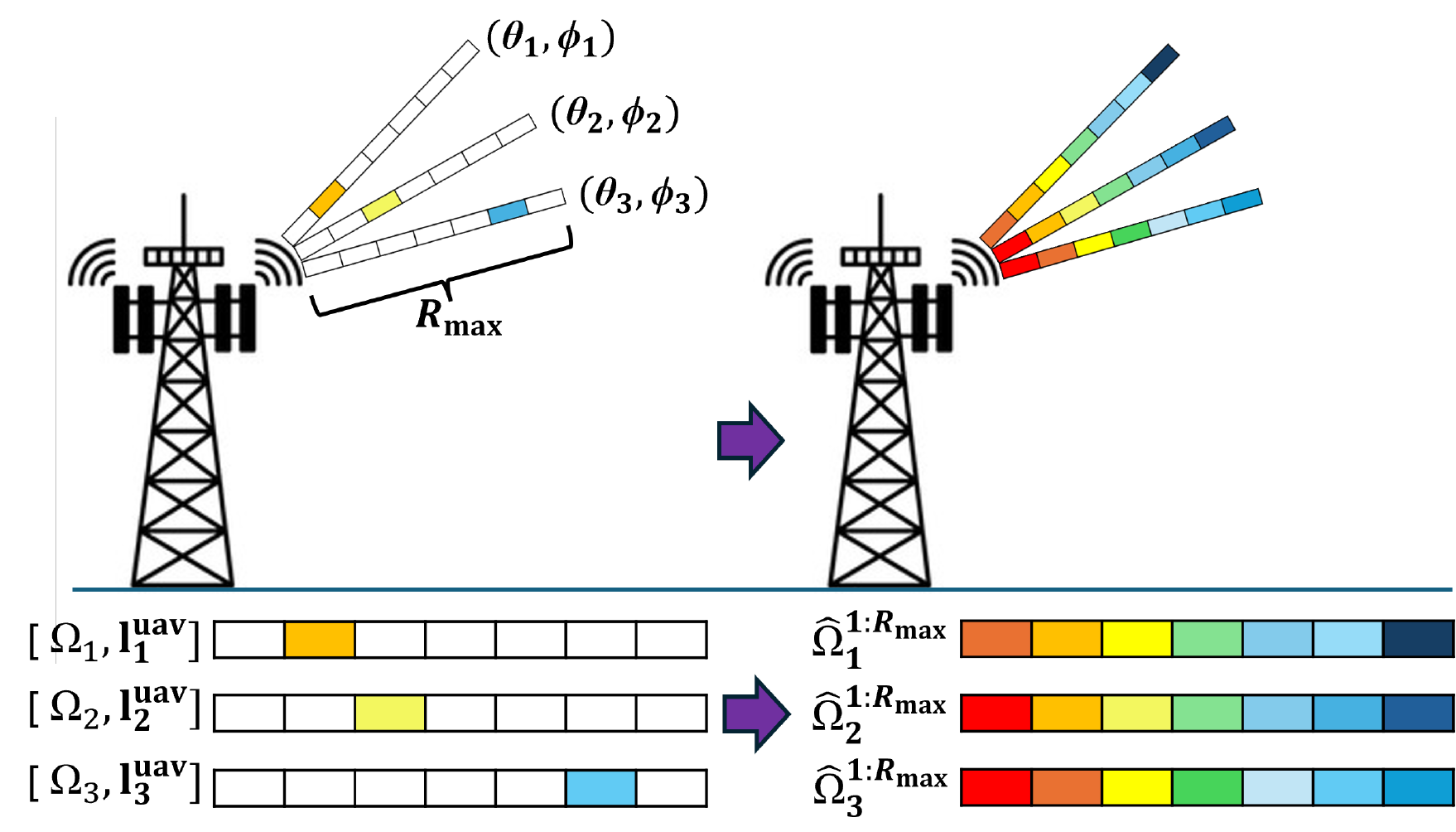}
	\centering
	\caption{We formulate 3D \gls{rem} generation through a 1D \gls{rsrp} sequence prediction task, exploiting the dominant \gls{los} propagation characteristics. Each measurement $[\Omega_i,\mathbf{l}_i^{\rm uav}]$ along the direction $(\theta_i, \phi_i)$ is used to reconstruct the full 1D sequence based on the underlying radial propagation profile.}
    \label{Fig:3DREMlearning}
\end{figure}

\begin{figure*}
	\includegraphics[width=0.9\linewidth]{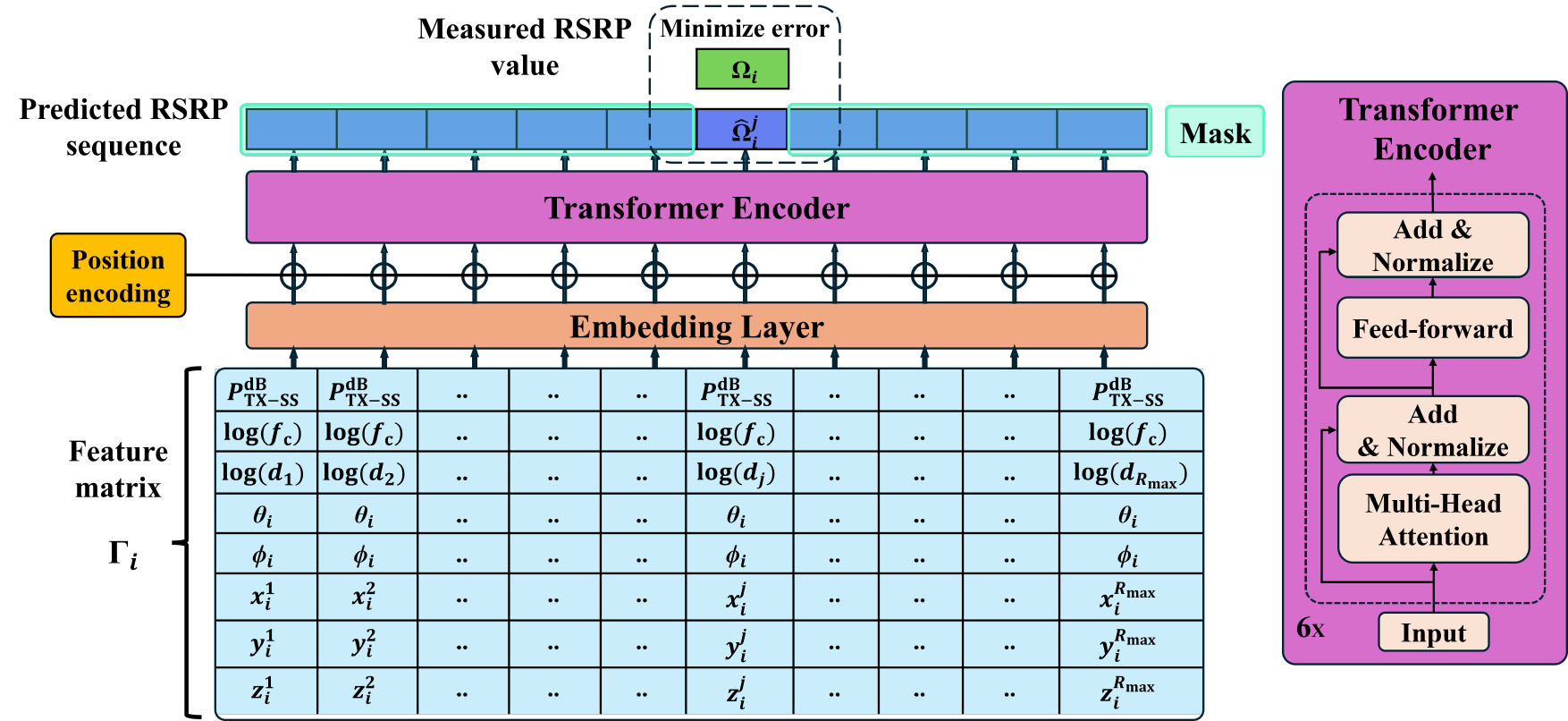}
	\centering
	\caption{For each measurement $i \in [1,...,N]$ along its direction ($\theta_i,\phi_i$) in 3D space, the input feature matrix $\mathbf{\Gamma}_i$ contains the large-scale \gls{los} channel parameters from~(\ref{eq:2ray_loss_split_dB}) and the transmit power $P_{\rm{TX-SS}}^{\rm{dB}}$ from~(\ref{eq:rsrp_power}) at each radial step from $d_1:d_{R_{\rm max}}$. The encoder combines the elements in $\mathbf{\Gamma}_i$ with attention weights to predict the \gls{rsrp} sequence $\widehat\Omega_i^{1:R_{\max}}$ for every step along this radial direction. During training, the predicted sequence is masked to isolate the specific radial bin $j$, corresponding to the ground truth \gls{rsrp} measurement $\Omega_i$, where $d_{\rm 3D}(\mathbf{l}^{\mathrm{bs}}, \mathbf{l}^{\mathrm{uav}}_i) \simeq d_j$. Subsequently, the loss between $\widehat\Omega_i^j$ and $\Omega_i$ is backpropagated for weight updates. The trained encoder can then generate complete 3D \glspl{rem} from \gls{rsrp} sequences over all ($\theta, \phi$).}   
    \label{fig:TransformerModel}
\end{figure*}

\subsection{Implications for Deep Learning Architecture}

The classical propagation models and empirical data analysis presented in this section serve as the physical blueprint for our proposed deep learning framework. First, the mathematical distinction between the \gls{los} decay and the stochastic multipath deviations highlights a fundamental challenge. Monolithic spatial models cannot easily learn both phenomena simultaneously from sparse real-world data.

Furthermore, our empirical data analysis reveals strong radial spatial correlation and sequential continuity in temporal fast-fading values.
Unlike classical interpolation methods such as 3D Kriging, which rely on generic multi-directional spatiograms~\cite{SungJoonKriging}.
This empirical finding explicitly justifies our departure from traditional 2D/3D grid-based models, motivating the novel grid-less, 1D radial sequence design of our proposed system with physically decoupled architecture and stage-wise training.


\section{REM SYSTEM DESIGN}\label{sec:ModelArchitecture}

In our proposed architecture, a spatial Transformer model first learns the \gls{los} path loss profile, followed by a \gls{gru} head to estimate sequential fast-fading behavior based on recent \gls{rsrp} data.
This hierarchical approach is then realized through a multi-stage training process described below.

\subsection{Transformer-based Spatial Modeling}

To construct 3D \glspl{rem} from sparse volumetric \gls{rsrp} measurements, we leverage the predominance of \gls{los} propagation in aerial scenarios.
Motivated by the strong radial correlation of signal strength, we formulate \gls{rem} construction as a sequence completion task for radio paths originating from the \gls{bs}, seen in Fig.~\ref{Fig:3DREMlearning}.
This approach involves fitting \gls{los} path loss exponents and learning the antenna patterns defined in~(\ref{eq:2ray_loss_split_dB}) and (\ref{eq:rsrp_power}) to match the observed data in 3D angular space.
Specifically, we sample the physical channel parameters along each measurement direction at discrete $1$~m steps up to a maximum range $d_{R_{\max}}$, and then employ a Transformer encoder to map these parameters into a \gls{rsrp} sequence of length $R_{\max}$. 
This spherical framework facilitates the generation of high-fidelity 3D \glspl{rem} with quantized radial resolution while preserving the continuity of the angular signal distribution.

From a dataset $\mathcal{S}$ with \gls{rsrp} observations $[\Omega_i, \mathbf{l}_i^{\rm uav}]$ at $N$ 3D spatial positions, we formulate the input feature matrix~$\mathbf{\Gamma}_i$ for each point $i \in [1,...,N]$ along its corresponding radial direction $(\theta_i,\phi_i)$.
The feature matrix~$\mathbf{\Gamma}_i$ is an $8\times R_{\rm{max}}$ array, defined as:\newline
    $\mathbf{\Gamma}_i = $
$\begin{bmatrix}
    P_{\mathrm{TX-SS}}^{\rm dB} & ... & P_{\mathrm{TX-SS}}^{\rm dB} & .... & P_{\mathrm{TX-SS}}^{\rm dB} \\
    \log_{10}(f_{\rm c}) & ... & \log_{10}(f_{\rm c}) & ... & \log_{10}(f_{\rm c}) \\
    \log_{10}(d_1) & ... & \log_{10}(d_j) & ... & \log_{10}(d_{R_{\rm{max}}}) \\
    \theta_i & ... & \theta_i & ... & \theta_i \\
    \phi_i & ... & \phi_i & ... & \phi_i \\
    x_i^1 & ... & x_i^j & ... & x_i^{R_{\rm{max}}} \\
    y_i^1 & ... & y_i^j & ... & y_i^{R_{\rm{max}}} \\
    z_i^1 & ... & z_i^j & ... & z_i^{R_{\rm{max}}} \\
\end{bmatrix},$
\newline

where $P_{\mathrm{TX-SS}}^{\rm dB}$ is from~(\ref{eq:rsrp_power}), $\log_{10}(f_{\rm c})$, $\log_{10}(d_j)$, $\theta_i$, and $\phi_i$ are the physical channel parameters at each radial step $j$ from~(\ref{eq:2ray_loss_split_dB}), and $x_i^{j}$, $y_i^j$, and $z_i^j$ are the cartesian coordinates at~$j^{\rm th}$ radial bin position along the $i^{\rm th}$ point's radial direction.
The input vector $\Gamma_i$ is intentionally restricted to observable macroscopic parameters. These parameters are mathematically sufficient to bound the large-scale channel attenuation, but fundamentally insufficient to capture fast-fading effects.

Our objective is to map underlying propagation conditions to a continuous \gls{rsrp} sequence $\widehat\Omega_i^{1:R_{\max}}$, that aligns with the observed value $\Omega_i$ at the radial bin $j$, where $d_{\rm 3D}(\mathbf{l}^{\mathrm{bs}}, \mathbf{l}^{\mathrm{uav}}_i) = d_j$. To capture these complex sequences, we employ an encoder-only Transformer architecture, as shown in Fig.~\ref{fig:TransformerModel}, adapted from the foundational wireless model framework~\cite{largewirelessmodel}. The transformer encoder processes the feature matrix $\mathbf{\Gamma}_i$ and utilizes attention mechanisms to combine the physical channel parameters at each radial bin into corresponding \gls{rsrp} sequences. 
The input features are initially projected through an embedding layer to map them into a higher-dimensional manifold, followed by the application of positional encodings to capture vital sequential dependencies.
The backbone of the model comprises six stacked encoder layers, each integrating a multi-head attention mechanism configured with eight heads.
The attention heads individually apply attention weights to aggregate the higher-dimensional input features across the length of the sequence.
This generates a predicted \gls{rsrp} profile that models signal attenuation along each angular direction, governed by the underlying channel parameters. 

\subsection{GRU-based Fast-Fading Factor Estimation}

Unlike free-space path loss, small-scale fast-fading is driven by undetermined micro-environmental variables such as the material properties of nearby scatterers, instantaneous UAV orientation, and dynamic blockages. Because a global 3D CAD model of every scatterer is computationally prohibitive to feed into the spatial encoder, fast-fading is inherently undetermined from the spatial coordinates alone. To resolve this, we leverage the GRU to temporally extrapolate these hidden multi-path contributors by observing the continuity of the signal over the \gls{uav}'s recent trajectory~\cite{TemporalModeling}.

\begin{figure}
	\includegraphics[width=0.9\linewidth]{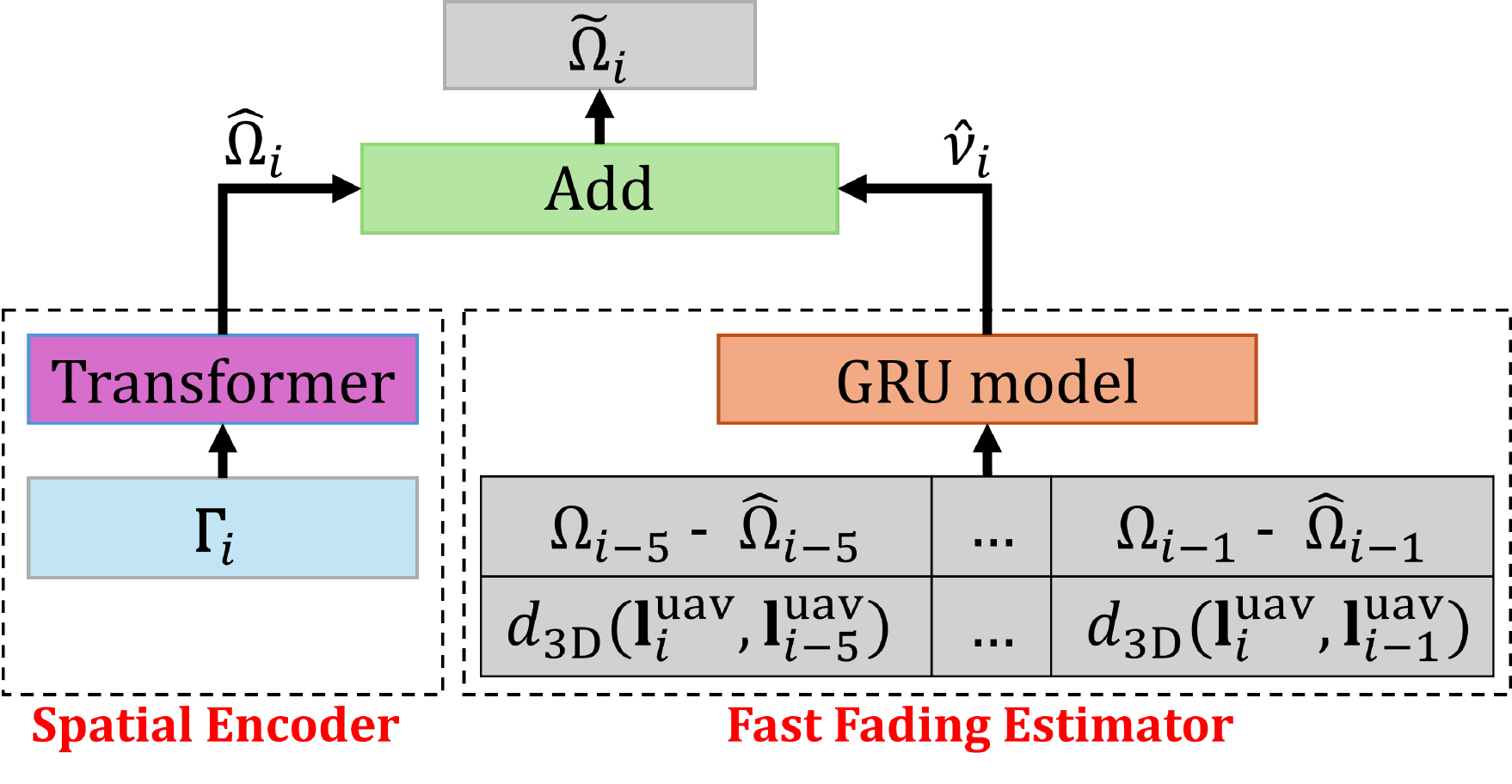}
	\centering
	\caption{The spatial encoder creates a global spatial \gls{rem} while the fast-fading estimator improves \gls{rsrp} predictions based on the recent fast-fading factors.}
    \label{Fig:arch}
\end{figure}

Fig.~\ref{Fig:arch} illustrates the combined architecture of the spatial and sequential \gls{rsrp} predictor model. 
Upon predicting the spatial \gls{rsrp} estimate $\widehat\Omega_i$ at a given position, the \gls{gru} sequential estimator updates the final estimate by adding a fast-fading prediction value $\widehat\nu_i$.
The sequential estimator takes the fast-fading factors from the previous five samples, calculated as $\nu_{i-m} = \Omega_{i-m} - \widehat\Omega_{i-m}$, along with their relative separation distance $d_{\rm{3D}}(\mathbf{l}^{\mathrm{uav}}_i, \mathbf{l}^{\mathrm{uav}}_{i-m})$ for $m \in \{1,...5\}$, to predict the current fast-fading value $\widehat\nu_i$ at position $i$.
The fast-fading estimate is combined with the spatial prediction to provide the fused \gls{rsrp} output $\widetilde\Omega_i$.
The temporal sequence length for the \gls{gru} is set to $T=5$ historical samples. This specific window size is chosen to optimize the trade-off between capturing sufficient temporal context and adhering to the physical limits of channel coherence. Due to the high mobility of the UAV, the localized scattering environment decorrelates rapidly over distance, requiring the relative separation between the 5 samples as the secondary input.

\subsection{Three-Stage Training Technique}\label{subsec:StagewiseTraining}

The combined \gls{rem} architecture in Fig.~\ref{Fig:arch} is trained in a three-stage manner designed to learn the physical channel effects progressively. This step-wise training procedure allows the transformer model to create a static spatial map with large-scale channel effects, followed by the \gls{gru} to estimate the residual fast-fading values, as follows:
\begin{enumerate}
   \item \textbf{Stage-1: Spatial encoder pre-training.} The spatial encoder is pre-trained using synthetic \gls{rsrp} data generated using the \gls{los} path loss model~(\ref{eq:2ray_loss_split_dB}) excluding BS antenna patterns. The pre-training on synthetic \gls{los} data acts as a geometric primer. It forces the multi-head attention weights to learn broad spatial position correlations and the underlying frequency and distance-dependent decay across the sequence length. The masked sequence training from Fig.~\ref{fig:TransformerModel} is performed with a \gls{mse} error minimization against the target synthetic \gls{rsrp} using a \gls{lwsrd} learning rate adaptation algorithm to optimally reach convergence.
    \item \textbf{Stage-2: Spatial encoder fine-tuning.} The pre-trained spatial encoder is then fine-tuned with real-world data~(\ref{eq:rsrp_power}) to learn regional shadowing effects and also to learn antenna patterns as observed in the real world. The pretrained model acts as a regularizing anchor, allowing the model to adapt to complex, real-world fading behaviors without overfitting to the specific fast-fading signatures of the training data. The model fine-tuning is performed with a smooth $\rm{\ell}_1$-loss function to promote robustness against fluctuations in the real-world data. 
    \item \textbf{Stage-3: Sequential encoder training.} The pre-trained spatial encoder is frozen for this stage of training. The \gls{gru} model alone is trained with a smooth $\rm{\ell}_1$-loss function to estimate the fast-fading factors and improve the final \gls{rsrp} estimate.  
\end{enumerate}

Our proposed training strategy helps to reconstruct continuous 3D \glspl{rem} from sparse, 2D \gls{uav} flight trajectories despite its inherently underdetermined nature. Conventional 3D grid-based networks struggle with this extreme sparsity, as unvisited voxels lack direct supervision. Our radial sequence Transformer efficiently mitigates this through the global receptive field of its self-attention mechanism. Although a real-world \gls{rsrp} measurement provides supervision for only a single coordinate within the $R_{\rm{max}}$-dimensional radial sequence, the error gradient backpropagates through the entire dense attention matrix. Because every token attends to all other tokens, a single sparse measurement dynamically calibrates the spatial embeddings and attention weights for the entire propagation ray. Furthermore, the Stage-1 synthetic pre-training serves as a critical structural prior. By anchoring the attention weights to the deterministic inverse-square law, it constrains the otherwise underdetermined 3D problem, ensuring that the model's spatial extrapolation into unmeasured airspace remains physically bounded rather than arbitrarily interpolating between sparse flight trajectories.
In this manner, unlike purely data-driven deep learning models, which struggle with extreme data sparsity and out-of-distribution extrapolation in 3D environments, our framework adopts a physics-informed learning paradigm.

\subsection{Baseline Methods for Comparison}\label{subsec:BaselineModels}

In this section, we describe the baseline methods used for benchmarking and introduce their architectures. We select several widely used architectures, including UNet~\cite{unet, levie2021radiounet}, Mamba~\cite{gu2024mamba}, Inception~\cite{ismail2020inceptiontime}, and 3D Kriging~\cite{SungJoonKriging} to compare with our proposed method. As the main goal is to provide a comprehensive comparison of different architectures, these baselines cover convolutional encoder-decoder, state-space, multi-scale convolutional architectures and a classical statistical technique. To maintain fairness across models, all models employ the same data preprocessing pipeline, loss function, and optimizer. Only the network architecture differs across experiments.

\subsubsection{UNet}
The first baseline architecture is UNet, which is a \gls{cnn} based on an encoder-decoder structure with symmetric skip connections. UNet has been known for its strong performance in dense prediction and masking tasks because it can capture both global context through downsampling and fine-grained local features through skip connections~\cite{unet}. 

RadioUNet~\cite{levie2021radiounet} extends the original UNet architecture to \gls{rem} construction from city-scale street maps. It learns spatial correlations from high-resolution 2D ray-tracing data. However, our approach differs from RadioUNet in that \gls{rem} is recreated from sparse one-dimensional\footnote{We refer to the dimension of the input data, not the propagation environment.} measurements, rather than 2D inputs of city maps.
Our approach differs from the state-of-the-art works in the literature. They mainly utilize 3D data (2D images of environments with signal strength), which creates very sparse representations in aerial networks. Therefore, posing 3D REM generation as a 1D radial sequence task is a deliberate design choice driven by the unique constraints of aerial networks. In \gls{uav} networks, the environment is fundamentally 3D and highly dynamic, often lacking the dense and static 2D environmental building maps that the original RadioUNet requires to work. Furthermore, relying on 3D volumetric convolutions for sparse aerial measurements introduces severe computational overhead and degradation due to sparsity~\cite{choy20194d, liu2019point}. Because this kind of method seems inappropriate for aerial systems, we adopt a UNet having a similar architecture to RadioUNet, but with one-dimensional input. To do so, we replace all 2D convolutions, pooling, and upsampling operations with their 1D counterparts. The resulting encoder consists of four downsampling stages with two convolutional layers followed by batch normalization and ReLU activation, and max-pooling with a stride of two. A bottleneck block further increases the receptive field. Then, a symmetric decoder upsamples the feature maps step-by-step using transposed convolutions. Also, we pad the input sequence to ensure divisibility by the total downsampling factor and obtain its original length at the output. This architecture produces a sequence-to-sequence prediction by the masking strategy detailed in the previous section.  

\begin{table}
    \centering
    \caption{Collection of three real-world datasets and their data splits.}
    \begin{tabular}{c|c|c|c}
         &  Nemo data & PawPrints data & Quectel data\\
    \hline
    Training samples  & 4029 & 2538 & 2822\\
    Test samples & 644 & 644 & 2792 \\
    \hline 
    Total samples & 4673 & 3182 & 5614 \\
    \end{tabular}
    \label{tab:datasetsummary}
\end{table}

\subsubsection{Mamba}
In \cite{gu2024mamba}, Mamba has been recently proposed as an efficient alternative to transformers and convolutional networks for long-sequence processing. It is based on a selective \gls{ssm}~\cite{gu2021efficiently}. While transformers' self-attention mechanism scales quadratically, Mamba performs with linear complexity. Also, it can model long-range dependencies through a recurrent state-space process. Another promising aspect is that Mamba's global memory evolves over the sequence, unlike \gls{cnn}'s fixed receptive field. 

Our benchmark model consists of four stacked Mamba blocks. In each block, there are pre-normalization layer, a gated linear unit, a depthwise convolution for local feature mixing, and a selective state-space scan, respectively. The state-space parameters are learned end-to-end. Therefore, adaptive modeling of both short-term variations and long-range propagation effects can be possible. As usual, the model has input and output projection operations that map the channel parameters to estimate \gls{rem} points.

\begin{table*}[t]
\centering
\caption{Performance results of the proposed models\textsuperscript{\textdagger} vs. the baseline techniques across the three datasets. Arrows indicate direction of improvement ($\downarrow$ lower is better, $\uparrow$ higher is better).}
\label{tab:results}
\resizebox{\textwidth}{!}{%
\begin{tabular}{lccccccccccc}
\toprule
\multirow{2}{*}{\textbf{Model}} & \multicolumn{3}{c}{\textbf{Nemo Dataset}} & \multicolumn{3}{c}{\textbf{PawPrints Dataset}} & \multicolumn{3}{c}{\textbf{Quectel Dataset}} & \textbf{Training} \\
& \multicolumn{3}{c}{} & \multicolumn{3}{c}{} & \multicolumn{3}{c}{(Train: Yaw45, Test: Yaw315)} & \textbf{Parameters}\\
\cmidrule(lr){2-4} \cmidrule(lr){5-7} \cmidrule(lr){8-10}
& RMSE $\downarrow$ & MAE $\downarrow$ & $R^2 \uparrow$ & RMSE $\downarrow$ & MAE $\downarrow$ & $R^2 \uparrow$ & RMSE $\downarrow$ & MAE $\downarrow$ & $R^2 \uparrow$ & \\
\midrule
\gls{los} PL model & 6.21 & 4.96 & -1.09 & 5.57 & 4.49 & -0.27 & 6.02 & 4.75 & -0.09 & --\\
3D Kriging & 2.73 & 2.23 & 0.59 & 3.20 & 2.55 & 0.57 & 4.95 & 3.89 & 0.26 & --\\
\midrule
Transformer\textsuperscript{\textdagger}  & 3.00 & 2.39 & 0.51  & 3.23 & 2.61 & 0.57  & 4.79 & 3.80 & 0.30  & $1{,}687k$ \\
Transformer+\gls{gru}\textsuperscript{\textdagger}  & \textbf{2.52} & \textbf{1.93} & \textbf{0.65} & 3.04 & 2.48 & 0.61 & \textbf{3.35} & \textbf{2.55} & \textbf{0.66} & $1{,}697k$ \\
\midrule
UNet                 & 3.60 & 2.85 & 0.47  & \textbf{2.52} & \textbf{1.95} & \textbf{0.83}  & 14.95 & 5.85 & -5.74 & $1{,}082k$ \\
Mamba                & 3.41 & 2.67 & 0.53  & 2.56 & 1.95 & 0.82  & 4.72  & 3.72 & 0.32  & $602k$ \\
Inception            & 3.60 & 2.80 & 0.47  & 2.82 & 2.10 & 0.78  & 8.94  & 4.73 & -1.41 & $218k$ \\
\bottomrule
\end{tabular}%
}
\end{table*}

\subsubsection{Inception}
Another baseline model is the Inception architecture, which was originally proposed to address the kernel-size selection problem in \gls{cnn}~\cite{szegedy2015going}. Inception modules apply multiple convolutional filters with different kernel sizes in parallel, and therefore, the network can capture features at multiple spatial or temporal scales simultaneously. As mentioned earlier, \gls{cnn} is limited by a constant receptive field. Inception architecture deals with this limitation. 

As our REM reconstruction task relies on one-dimensional input sequences, we adopt InceptionTime~\cite{ismail2020inceptiontime}, which employs the Inception architecture for time-series regression and classification tasks. Each Inception module consists of a bottleneck layer that reduces channel dimensionality, followed by parallel 1D convolutions with multiple kernel sizes and an additional max-pooling branch. Each output is concatenated and normalized before being passed to the next layer. A final pointwise convolution maps the aggregated features to a single REM value.

By stacking multiple InceptionTime blocks, we believe that the model can learn hierarchical multi-scale representations of the input signal, which might be helpful for \gls{rem} reconstruction, where measurements are under the impacts of both small-scale fading and large-scale shadowing phenomena occurring over different spatial scopes.

\subsubsection{3D Kriging}
Finally, to evaluate against classical statistical methods, we employ 3D Kriging~\cite{SungJoonKriging}. Kriging is a widely adopted geostatistical interpolation method that is traditionally considered a standard method for \gls{rem} construction. It estimates the signal strength at unmeasured locations through a distance-weighted linear combination of known sparse measurements. The optimal weights are mathematically derived from a spatial correlation model, known as a semivariogram, which characterizes how the variance of the radio signal changes with spatial separation.

Following the high-level implementation~\cite{SungJoonKriging}, we utilize 3D Ordinary Kriging. In this 3D adaptation, the empirical semivariogram is computed directly over the three-dimensional spatial coordinates to capture the volumetric spatial correlation and distance-dependent decay of the aerial measurements. This allows the model to interpolate the continuous radio map across different flight altitudes and spatial planes based on structural, distance-based dependencies. While 3D Kriging is highly effective for reconstructing static spatial maps and serves as a robust baseline, it relies exclusively on generalized multi-directional spatiograms.


\section{EVALUATION AND RESULTS}
\label{sec:Results}

In this section, we evaluate the \gls{rem} construction models and compare their relative performance.
To test the models, we used the three datasets from Section~\ref{sec:SystemModel}-\ref{SubSec:Datasets}, each treated as separate training and testing datasets. 
The test portions were selected over continuous time intervals to ensure the complete isolation of the training and testing sets.
The dataset splits are detailed in Table~\ref{tab:datasetsummary}.
The Nemo and PawPrints \gls{rsrp} measurements were collected at altitudes of 30~m and 50~m above the ground, using the same \gls{uav} trajectory relative to the \gls{bs}.
For the Quectel dataset alone, we use the entire $45^{\circ}$ yaw angle data for training and the $315^{\circ}$ yaw angle data for testing.

Utilizing the three datasets, we train the spatial encoder and sequential estimator as per the stagewise training procedure detailed in Section~\ref{sec:ModelArchitecture}-\ref{subsec:StagewiseTraining}.
Similarly, the baseline models detailed in Section~\ref{sec:ModelArchitecture}-\ref{subsec:BaselineModels} are also trained to predict \gls{rsrp} values in the test set. 
To evaluate prediction accuracy, we employ the \gls{rmse}, the \gls{mae}, and the coefficient of determination ($R^2$) to assess the model's ability to capture the statistical variability in the datasets.
The $R^2$ coefficient is defined as: 
\begin{equation}
    \begin{aligned}
        R^2 &= 1 - \frac{\sum_{i=1}^{N}( \Omega_i - \hat{\Omega}_i )^2}{\sum_{i=1}^{N}( \Omega_i - \bar{\Omega} )^2},
    \end{aligned}
\end{equation}
where $\Omega_i$ is the real value, $\hat{\Omega}_i$ is the predicted value, and $\bar{\Omega}$ is the average of the $N$ real values in the test set.

\begin{figure*}
	\includegraphics[width=0.8\linewidth]{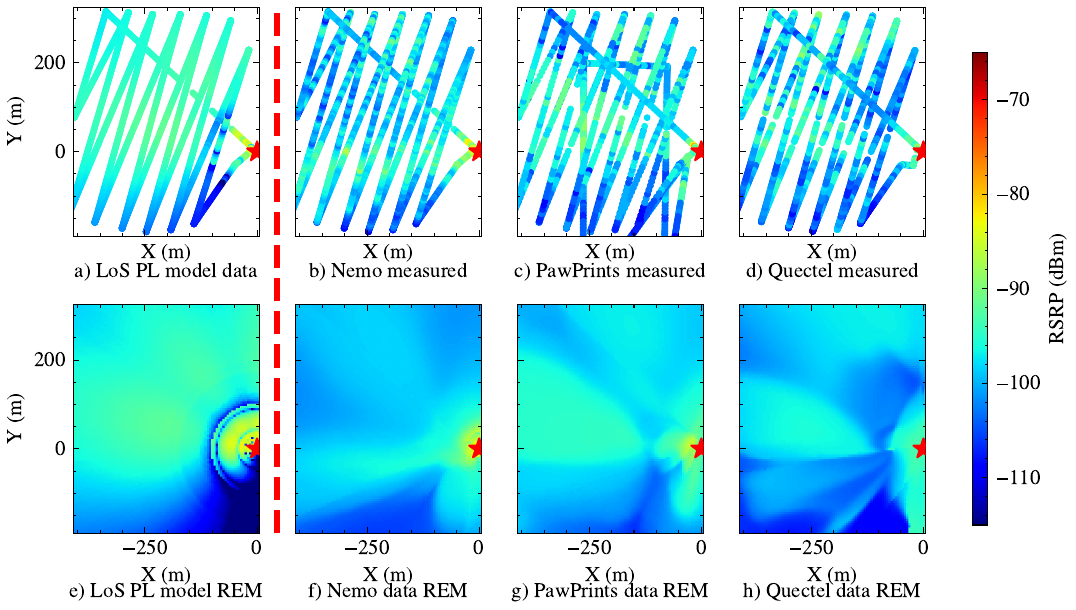}
	\centering
	\caption{Data measured at 30~m altitude from the various \gls{uav} mounted receivers with respect to the \gls{bs} at origin (Quectel yaw $45^{\circ}$ data). The corresponding predicted heatmaps using the proposed transformer spatial model are seen in subplots f), g), and h). Subplots a) and e) illustrate the synthetic data and spatial heatmap generated from the deterministic \gls{los} path loss model~(\ref{eq:2ray_loss_split_dB}) in the same spatial coordinates for comparison.}
    \label{Fig:REMheatmaps}
\end{figure*}

\subsection{RSRP Prediction Performance}

To evaluate sensitivity to measurement precision on the \gls{ml} models, we employ the Nemo and PawPrints datasets with intra-distribution testing. This allows for a critical comparison between Nemo’s high-resolution data ($0.5$~s intervals) and PawPrints’ integer-quantized reports ($1$~s intervals). In contrast, the Quectel dataset ($0.25$~s interval) is reserved for evaluating generalization across \gls{uav} orientation dynamics, utilizing a test set distinct from the training distribution.
Table~\ref{tab:results} contains the evaluation results, beginning with a deterministic \gls{los} path loss model-based fit showing high errors.
3D Kriging is highly effective. It significantly outperforms the baseline \gls{los} path loss model across all datasets. On the Nemo dataset, it outperforms the deep learning baselines, achieving under $3$~dB \glspl{rmse} and under $5$~dB \gls{rmse} for the out-of-distribution Quectel test set.
The proposed spatial transformer encoder mapping the large-scale fading learns separate \glspl{rem} from all three datasets, showing test set \gls{rmse} values below $5$~dB, and $R^2$ values above $0.30$. 
The Quectel model shows the lowest accuracy due to the different fast-fading pattern in the test set, driven by a changed receiver orientation over the same trajectory.
The combined transformer+\gls{gru} model improves on these results by estimating the fast-fading effects from recent samples, thus bringing the \gls{rmse} values close to $3$~dB and increasing the $R^2$ values above $0.60$ across all datasets.
It is interesting to note that the Quectel and Nemo benefit more from the \gls{gru} estimation, owing to the better correlation of fast-fading values observed in Section~\ref{sec:SystemModel}-\ref{subsec:CorrelationAnalysis}

Among the baseline models, the Mamba architecture is the most consistent, with results nearly the same as those of the spatial transformer encoder. 
Moreover, the baseline methods outperform the proposed models exclusively in the PawPrints dataset. This provides insight into the operational strengths of these architectures and can be explained by the relatively strong large-scale correlation and smoothly quantized fast-fading behavior observed in the PawPrints dataset. Because the PawPrints data was collected with a 1.0~s reporting interval and integer-quantized values, it inherently smooths out fine-grained fast-fading effects.
Convolutional and \glspl{ssm} tend to show good performance, as their characteristics favor smooth and slowly varying signal patterns.
For example, convolutional architectures show a strong inductive bias towards locality and translation invariance~\cite{kayhan2020, wang2024theoretical}; thus, they efficiently capture the smooth and spatially correlated features of large-scale fading.
Also, Mamba is inherently good at modeling the slowly varying dependencies as \glspl{ssm} are based on continuous-time control systems.
As reported in~\cite{ma2025rethinking}, \glspl{ssm} experience memory decay over sequence length, and they might fail to capture fast-changing characteristics.

These results highlight the specific necessity of the proposed Transformer+GRU architecture for realistic, high-mobility aerial links.  The baseline models struggle when deployed on the Nemo dataset, which consists of non-quantized, decimal-level measurements, or the Quectel dataset involving a rapid 0.25~s reporting period and out-of-distribution fast-fading dynamics induced by changed \gls{uav} orientation. This is most evident in UNet's performance degradation to an RMSE of 14.95~dB on the Quectel dataset. Standard convolutional and \glspl{ssm} fail to capture these rapid stochastic fluctuations. By decoupling the large-scale path loss, which is handled by the Transformer, from the highly dynamic, localized fast-fading, which is handled by the GRU, the proposed framework excels on dynamic datasets. The proposed method is capable of maintaining high-fidelity predictions with an \gls{rmse} of $3$~dB in environments where signal variability is rapid and unquantized, but yields diminishing returns on smoothed data.
Furthermore, the improved performance of the proposed \gls{rem} architecture is inherently tied to the increased model complexity. This is discussed in more detail in Subsection~\ref{sec:Results}-\ref{subsec:ModelComplexity}. 

\begin{figure*}
    \includegraphics[width=0.8\linewidth]{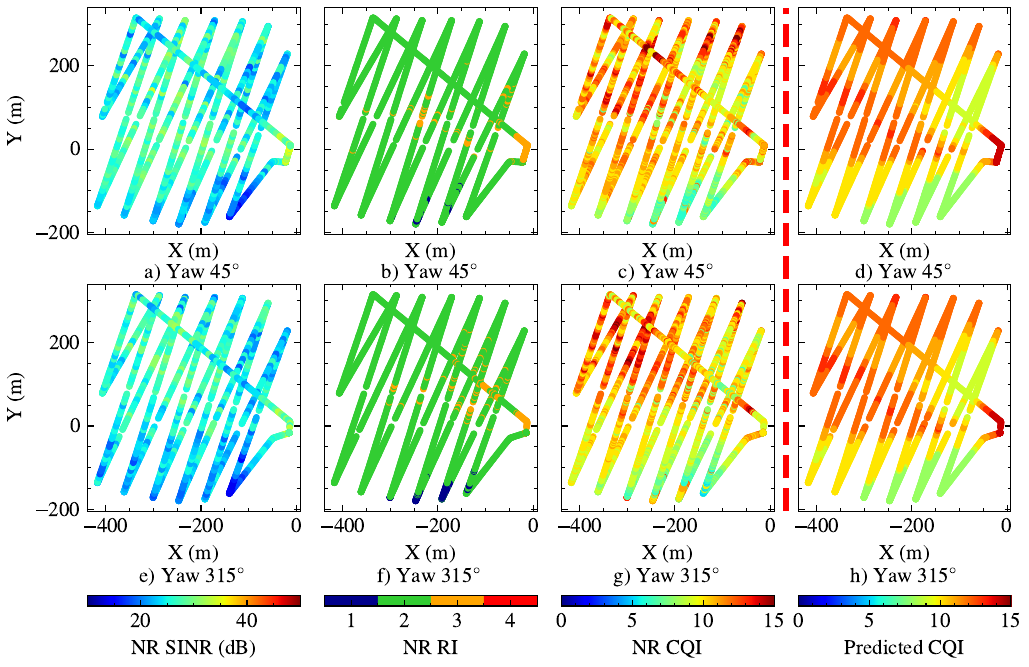}
	\centering
	\caption{The Quectel reported metrics from the yaw $45^{\circ}$ and yaw $315^{\circ}$ flights, with \glspl{cqi} in subplots c) and g) varying according to the observed \gls{sinr}  and \gls{ri} in subplots a), b) and e), f) respectively. The aerial channel around to the \gls{bs} is characterized by a dominant spatial rank of 2 and minimal interference. Consequently, d) and h) illustrate the large-scale \glspl{cqi} trends derived from the \gls{rem} shown in Fig.~\ref{Fig:REMheatmaps}d) for comparison.}
    \label{fig:QuectelCQI}
\end{figure*}

\subsection{Spatial REM Visualization}

To further validate the \gls{rem} learning capability of the proposed spatial transformer encoder, we visualize the generated maps at 30~m altitude surrounding the \gls{bs}, in Fig.~\ref{Fig:REMheatmaps}.
In the absence of comprehensive real-world heatmaps, we employ the deterministic \gls{los} path loss model heatmap as a proxy to approximate the idealistic large-scale fading characteristics in the \gls{bs} vicinity.
Next, the learned spatial transformer encoder generates \gls{rsrp} estimates at spatial intervals of $5\times5~\rm{m}^2$ in the horizontal plane, to visualize the learned \gls{rem}. 
The deterministic model-driven heatmap in Fig.~\ref{Fig:REMheatmaps}e) demonstrates the presence of upward-pointed sidelobes within a 100~m radius surrounding the \gls{bs}.
Beyond the 100~m radius region, the signal strength distribution is more gradual. 
The deterministic heatmap also demonstrates a poor coverage area in the bottom right of the subplot, which is due to the physical antenna tilt towards the northern region.

We assess the generated heatmaps against the deterministic \gls{los} model heatmap using the \gls{ssim} to evaluate the global \gls{rsrp} distribution and \gls{fsim} to measure the accuracy of antenna lobe patterns and coverage holes. These results are summarized in Table~\ref{tab:heatmapresults}.
Quantitatively, the PawPrints heatmap closely aligns with the deterministic model, illustrating the benefit of extended averaging intervals. However, it still exhibits discernible deviations that likely reflect underlying real-world propagation effects unknown to the idealistic heatmap.

\begin{table}
    \centering
    \caption{Similarity results of the predicted and deterministic heatmaps.}
    \begin{tabular}{c|c|c|c}
        Metric &  Nemo data & PawPrints data & Quectel data \\
        \hline
        \gls{ssim} & 0.75 & \textbf{0.80} & 0.73 \\
        \gls{fsim} & 0.86 & \textbf{0.86} & 0.85 \\
    \end{tabular}
    \label{tab:heatmapresults}
\end{table}

\subsection{Case Study: UAV Channel Quality Prediction}\label{sec:CaseStudy}

This case study is inspired by the prior work~\cite{UAV_CQI_Predictor}, which highlights the importance of accurate \gls{cqi} knowledge for \gls{uav} link reliability.
\gls{cqi} values quantify the channel quality at the receiver based on measured \gls{sinr}, represented as integers between 0 and 15. 
The Quectel data, containing \gls{sinr}, \gls{ri}, and \gls{cqi} logs, demonstrates that \gls{cqi} variation is governed by \gls{sinr} and \gls{ri}, as seen in Fig.~\ref{fig:QuectelCQI}. These measurements reveal the coverage area rich in rank 2 channels, with \gls{sinr} patterns similar to the \gls{rsrp} trends shown in Fig.~\ref{Fig:REMheatmaps}d).

\begin{table}
    \centering
    \caption{Top-$K$ CQI prediction accuracy on Quectel data. }
    \begin{tabular}{c|c|c|c}
         & Top-1 & Top-2 & Top-3 \\
        \hline
        Yaw $45^{\circ}$ & 26.25\% & 63.71\% & 85.35\% \\
        Yaw $315^{\circ}$ & 21.48\% & 59.45\% & 81.57\% \\
    \end{tabular}
    \label{tab:CQIpredictionresults}
\end{table}

\begin{table*}
\centering
\caption{Performance results of the proposed models\textsuperscript{\textdagger} when trained and tested with data from different altitudes. Arrows indicate direction of improvement ($\downarrow$ lower is better, $\uparrow$ higher is better).}
\label{tab:CrossAltitudeGeneralization}
\resizebox{\textwidth}{!}{%
\begin{tabular}{lcccccccc}
\toprule
\multirow{2}{*}{\textbf{Data Split}} & \multirow{2}{*}{\textbf{Model}} & \multicolumn{3}{c}{\textbf{Nemo Dataset}} & \multicolumn{3}{c}{\textbf{PawPrints Dataset}}\\
& & \multicolumn{3}{c}{} & \multicolumn{3}{c}{}\\
\cmidrule(lr){3-5} \cmidrule(lr){6-8}
& & RMSE (dB) $\downarrow$ & MAE (dB) $\downarrow$ & $R^2 \uparrow$ & RMSE (dB) $\downarrow$ & MAE (dB) $\downarrow$ & $R^2 \uparrow$\\
\midrule
            & Transformer\textsuperscript{\textdagger} & 5.25 & 4.37 & -0.10 & 5.77 & 4.79 & -1.03 \\
            & Transformer +\gls{gru}\textsuperscript{\textdagger} & \textbf{4.19} & \textbf{3.46} & \textbf{0.32} & 5.35 & 4.21 & -0.75 \\
Train: 30~m & 3D Kriging           & 5.55 & 4.69 & -0.16 & 6.02 & 5.07 & -1.17 \\
Test: 50~m  & UNet                 & 6.15 & 5.09 & -0.52 & \textbf{4.78} & \textbf{3.61} & 0.05 \\
            & Mamba                & 6.22 & 5.16 & -0.57 & 5.12 & 3.81 & \-0.05 \\
            & Inception            & 6.85 & 5.66 & -0.88 & 5.69 & 4.20 & -0.36 \\
\midrule
            & Transformer\textsuperscript{\textdagger} & 5.62 & 4.55 & -0.62 & 5.42 & 4.24 & -0.15 \\
            & Transformer +\gls{gru}\textsuperscript{\textdagger} & \textbf{5.15} & \textbf{4.07} & \textbf-0.32 & \textbf{4.87} & \textbf{3.91} & \textbf{0.06} \\
Train: 50~m & 3D Kriging           & 5.49 & 4.55 & -0.54 & 5.78 & 4.61 & -0.31\\
Test: 30~m  & UNet                 & 5.25 & 4.23 & \textbf{-0.13} & 5.60 & 4.40 & -0.31 \\
            & Mamba                & 5.25 & 4.21 & -0.15 & 5.97 & 4.66 & -0.48 \\
            & Inception            & 5.73 & 4.53 & -0.33 & 6.75 & 5.20 & -0.84 \\
\bottomrule
\end{tabular}%
}
\end{table*}

We extend the \gls{rem} in Fig.~\ref{Fig:REMheatmaps}h) into a \gls{sinr} map at $30$~m altitude, accounting for a constant interference-plus-noise of $-115$~dBm per \gls{nr} resource element. Subsequently, considering predominantly Rank 2 channels, consistent with the environment, we map these \gls{sinr} values with a look-up table to their corresponding \glspl{cqi}.  We then evaluate the predicted \gls{cqi} seen in Figs.~\ref{fig:QuectelCQI}d,~and~\ref{fig:QuectelCQI}h) against ground truth measurements. Here, we report the Top-$K$ accuracy based on an absolute error margin of $K$, considering a prediction successful if $|{\rm CQI}_{\rm pred} - {\rm CQI}_{\rm true}| \le {K}$. As detailed in Table~\ref{tab:CQIpredictionresults}, we achieve above $80$\% Top-3 accuracy, indicating that the \gls{rem} derived \gls{cqi} map aligns closely with observed channel quality, with most \gls{cqi} predictions falling within a margin of $\pm 3$ indices. This capability is instrumental in proactively managing future aerial links across 3D space.

\subsection{Cross-Altitude Generalization}

To validate the true 3D spatial mapping capabilities of the proposed architecture, we explicitly evaluate out-of-distribution cross-altitude generalization. The Nemo dataset comprises $3{,}033$ and $1{,}519$ measurements at the $30$~m and $50$~m altitudes, while the PawPrints dataset comprises $2{,}433$ and $643$ measurements at these altitudes, respectively. We train and test on these separate altitudes to observe such generalization.

Table~\ref{tab:CrossAltitudeGeneralization} details the performance of the proposed architecture against classical and deep learning baselines under these strict cross-altitude splits. Inherently, extrapolating \gls{rsrp} into unseen 3D spatial planes is highly challenging, resulting in broader error margins and reduced $R^2$ coefficients compared to mixed-altitude evaluations. Despite these challenges, the decoupled Transformer+\gls{gru} framework exhibits the highest robustness to altitude shifts among the evaluated baseline models. In the Nemo dataset, when trained at $30$~m and tested at $50$~m, the proposed model achieves an \gls{rmse} of $4.19$~dB, outperforming 3D Kriging and sequential models like Mamba. This demonstrates that the spatial Transformer effectively learns a robust underlying 3D macroscopic effect, enabling better zero-shot extrapolation than purely data-driven monolithic models. 

The baseline models like UNet and Mamba perform well on the PawPrints dataset when extrapolating from the $30$~m altitude to the $50$~m altitude, because the dataset has inherently smoothed fast-fading characteristics as explained in the previous sections. Also, the signal environment at $50$~m is relatively smooth, but it might become highly complex at $30$~m due to the impact of multiple antenna sidelobes nearby.  However, these models significantly struggle in the reverse case, because purely data-driven models cannot infer the unobserved, complex antenna sidelobes that might be present in \gls{rsrp} data of lower altitudes with sufficient detail. Hence, the proposed framework proves essential for bounding and extrapolating granular spatial \gls{rsrp} across 3D altitudes.

\subsection{Computational Complexity and System Trade-offs}
\label{subsec:ModelComplexity}

For highly dynamic \gls{uav} networks, \gls{rem} generation must satisfy strict real-time inference constraints to enable applications like proactive handovers. Table~\ref{tab:TransformerEncoderSizeVariation} presents a comprehensive evaluation of computational complexity, comparing parameter footprints and inference latencies across standard (NVIDIA T4) and advanced (NVIDIA G4) GPU architectures. The G4 GPU, based on the recent NVIDIA Blackwell architecture, is optimized for faster floating-point operation performance and serves to benchmark the framework on state-of-the-art hardware.

First, it is important to contextualize the differences in model size. The proposed optimal 6-layer Transformer+GRU model utilizes $1{,}697$k parameters, which is larger than the sequential Mamba baseline with $602$k parameters. In our Transformer architecture, the parameter count is determined exclusively by the network depth (number of encoder layers) and hidden dimensions, utilizing fixed sinusoidal positional encodings. Notably, it is completely independent of the spatial sequence length. Our hyperparameter evaluations revealed that a 6-layer architecture provides the optimal physical capacity necessary to decouple large-scale shadowing from fast-fading channel effects. The proposed model achieved \gls{rsrp} prediction \glspl{rmse} under $3$~dB even with a varying number of transformer encoder layers.

\begin{table*}
\centering
\caption{Performance results of the proposed models\textsuperscript{\textdagger} with varying number of transformer encoder layers and varying radial step size. Arrows indicate direction of improvement ($\downarrow$ lower is better, $\uparrow$ higher is better).}
\label{tab:TransformerEncoderSizeVariation}
\resizebox{\textwidth}{!}{%
\begin{tabular}{lccccccc}
\toprule
\multirow{2}{*}{\textbf{Model}} & \multicolumn{3}{c}{\textbf{Nemo Dataset}} & \textbf{Training} & \multicolumn{2}{c}{\textbf{Inference}}\\
& \multicolumn{3}{c}{\textbf{(Reporting Interval~$0.5$s)}} & \textbf{Parameters} & \multicolumn{2}{c}{\textbf{Time}}\\
\cmidrule(lr){2-4} \cmidrule(lr){5-7}
& RMSE (dB) $\downarrow$ & MAE (dB) $\downarrow$ & $R^2 \uparrow$ & & T4 GPU (ms) & G4 GPU (ms)\\
\midrule
Transformer (2 layers)+\gls{gru}\textsuperscript{\textdagger}  & 2.80 & 2.20 & 0.57 & $572$k & 0.71 & 0.11\\
Transformer (4 layers)+\gls{gru}\textsuperscript{\textdagger}  & 2.57 & 2.04 & 0.63 & $1{,}135$k & 1.19 & 0.17\\
Transformer (6 layers)+\gls{gru}\textsuperscript{\textdagger}  & \textbf{2.52} & \textbf{1.93} & \textbf{0.65} & $1{,}697$k & 1.76 & 0.22\\
Transformer (8 layers)+\gls{gru}\textsuperscript{\textdagger}  & 2.69 & 2.15 & 0.60 & $2{,}259$k & 2.15 & 0.28\\
\midrule
Transformer (6 layers, 1m radial step) +\gls{gru}\textsuperscript{\textdagger} & \textbf{2.52} & \textbf{1.93} & \textbf{0.65} & $1{,}697$k & 1.76 & 0.22\\
Transformer (6 layers, 5m radial step) +\gls{gru}\textsuperscript{\textdagger} & 2.65 & 2.13 & 0.61 & $1{,}697$k & 0.57 & 0.14\\
Transformer (6 layers, 10m radial step)+\gls{gru}\textsuperscript{\textdagger} & 2.76 & 2.23 & 0.58 & $1{,}697$k & 0.59 & 0.15\\
\midrule
3D Kriging                         & 2.73 & 2.23 & 0.59 & - & 1.93 & 0.51 \\
UNet                               & 3.60 & 2.85 & 0.47 & $1{,}082$k & 1.02 & 0.12 \\
Mamba                              & 3.41 & 2.67 & 0.53 &  $602$k & 0.95 & 0.13 \\
Inception                          & 3.60 & 2.80 & 0.47 &  $218$k & 0.29 & 0.04 \\
\bottomrule
\end{tabular}%
}
\end{table*}

Despite having a larger parameter footprint than Mamba, the proposed architecture is highly efficient. By varying the number of encoder layers, the T4 inference time ranges from $0.71$ to $1.76$~ms. This remains highly competitive with the Mamba baseline ($0.95$~ms) and executes faster than the classical 3D Kriging baseline ($1.93$~ms). These latencies further improve on the G4 GPU, demonstrating that the model's execution time is well within the typical reporting intervals of UAV networks.

While the parameter count is fixed, the true computational bottleneck of a Transformer lies in the $O(N^2)$ self-attention footprint. However, this characteristic provides network operators with a flexible trade-off between the sequence length ($N$) and inference time. By increasing the spatial interpolation step size ($\Delta R$) from $1$~m to $5$~m, the radial sequence length is drastically reduced. Because the attention mechanism scales quadratically, Table~\ref{tab:TransformerEncoderSizeVariation} demonstrates that this minor reduction in spatial granularity slashes the T4 inference time from $1.76$~ms down to $0.57$~ms. This only costs a marginal penalty to predictive accuracy, further underscoring the real-time viability of our proposed grid-less framework. This trade-off eventually saturates, depending on the total radial distance, as observed by the identical inference times between the $5$~m and $10$~m step sizes. Finally, the role of sequence length scaling distinctly impacts the baselines as well. While the Mamba model benefits from a linear $O(N)$ footprint, classical 3D Kriging suffers from an $O(N^3)$ computational footprint.


\glsresetall

\section{CONCLUSION}
\label{sec:Conclusion}
In this work, we propose a cascaded \gls{rem} framework that decouples the learning of large-scale spatial radio footprints from localized fast-fading dynamics.
By leveraging the inherent spatio-temporal correlation properties found in empirical datasets, the proposed architecture achieves high-fidelity 3D aerial signal strength predictions. 
We benchmark our framework against state-of-the-art sequence and state-modeling techniques, demonstrating consistent predictions with errors near $3$~dB and cross-dataset generalization to construct \glspl{rem} with similarity indexes above $0.75$.
Finally, we showcase the utility of the generated \gls{rem} in predicting \gls{uav} link quality, which is a critical capability for optimizing coordinated aerial network operations and trajectory planning. 

\section*{ACKNOWLEDGMENT}
We would like to thank Dr. Simran Singh and Dr. \"Ozg\"ur \"Ozdemir from NC State University for the dataset collection and valuable discussions.

\bibliographystyle{IEEEtran}
\bibliography{Bibtex/refs}

\begin{IEEEbiography}[{\includegraphics[width=1in,height=1.25in,clip,keepaspectratio]{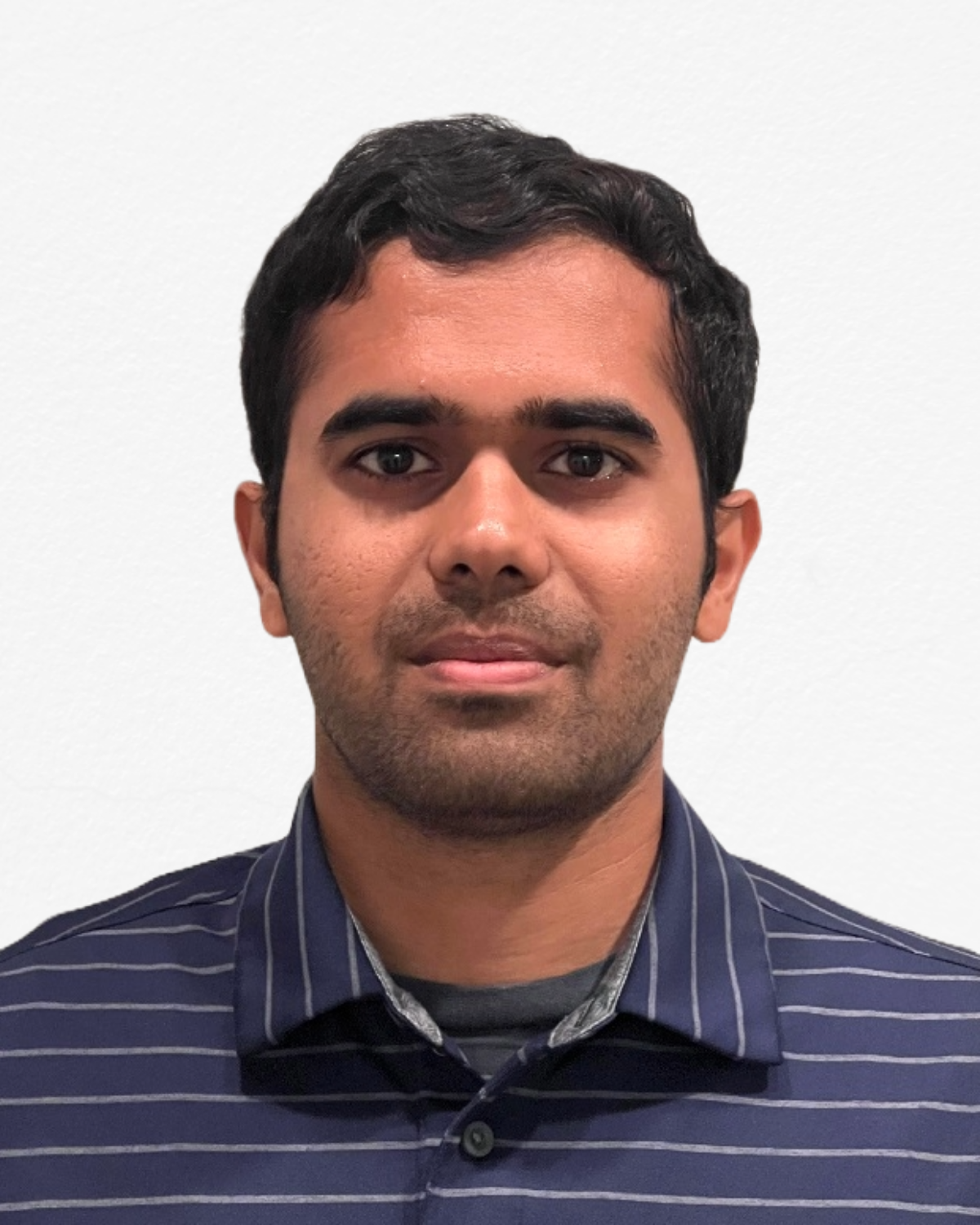}}]
{GAUTHAM REDDY}~received his M.Sc. in Communication Theory and Systems from U.C. San Diego in 2019. He then worked at Marvell Semiconductors as part of the Wireless Communication Algorithms team for two years. He is currently a graduate student pursuing his Ph.D. at North Carolina State University. His research interests include 5G, Next Gen communication system architecture, and xApp design in ORAN/AI-RAN networks. 
\end{IEEEbiography}

\begin{IEEEbiography}[{\includegraphics[width=1in,height=1.25in,clip,keepaspectratio]{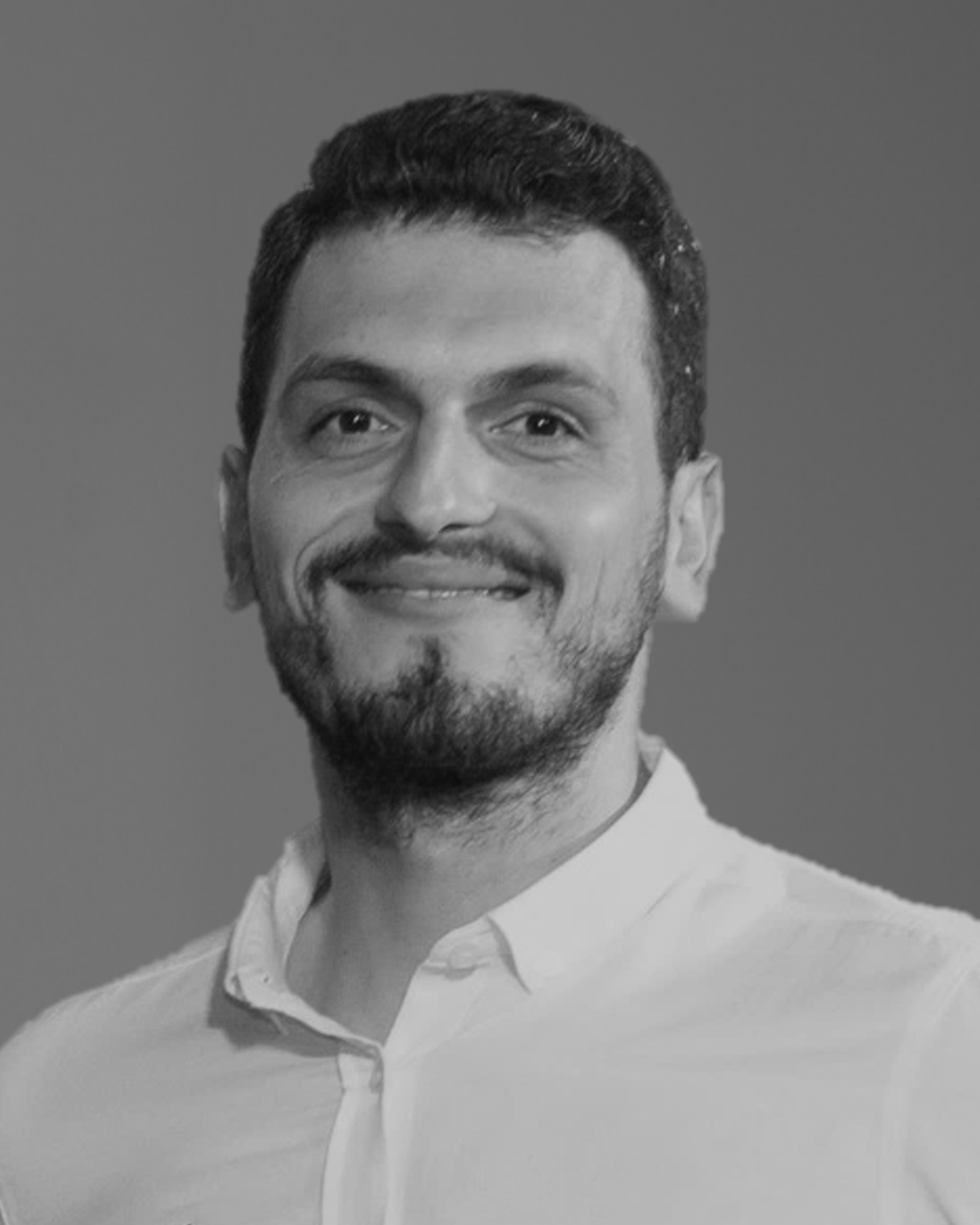}}]
{K\"{U}R\c{S}AT TEKBIYIK}~(Member,~IEEE) received his B.Sc., M.Sc., and Ph.D. degrees in Telecommunications Engineering from Istanbul Technical University, Turkey, in 2017, 2019, and 2024, respectively. His previous works cover areas ranging from signal intelligence systems to terahertz communications. His research interests include nonterrestrial networks and machine learning applications in wireless communications.
\end{IEEEbiography}

\begin{IEEEbiography}[{\includegraphics[width=1in,height=1.25in,clip,keepaspectratio]{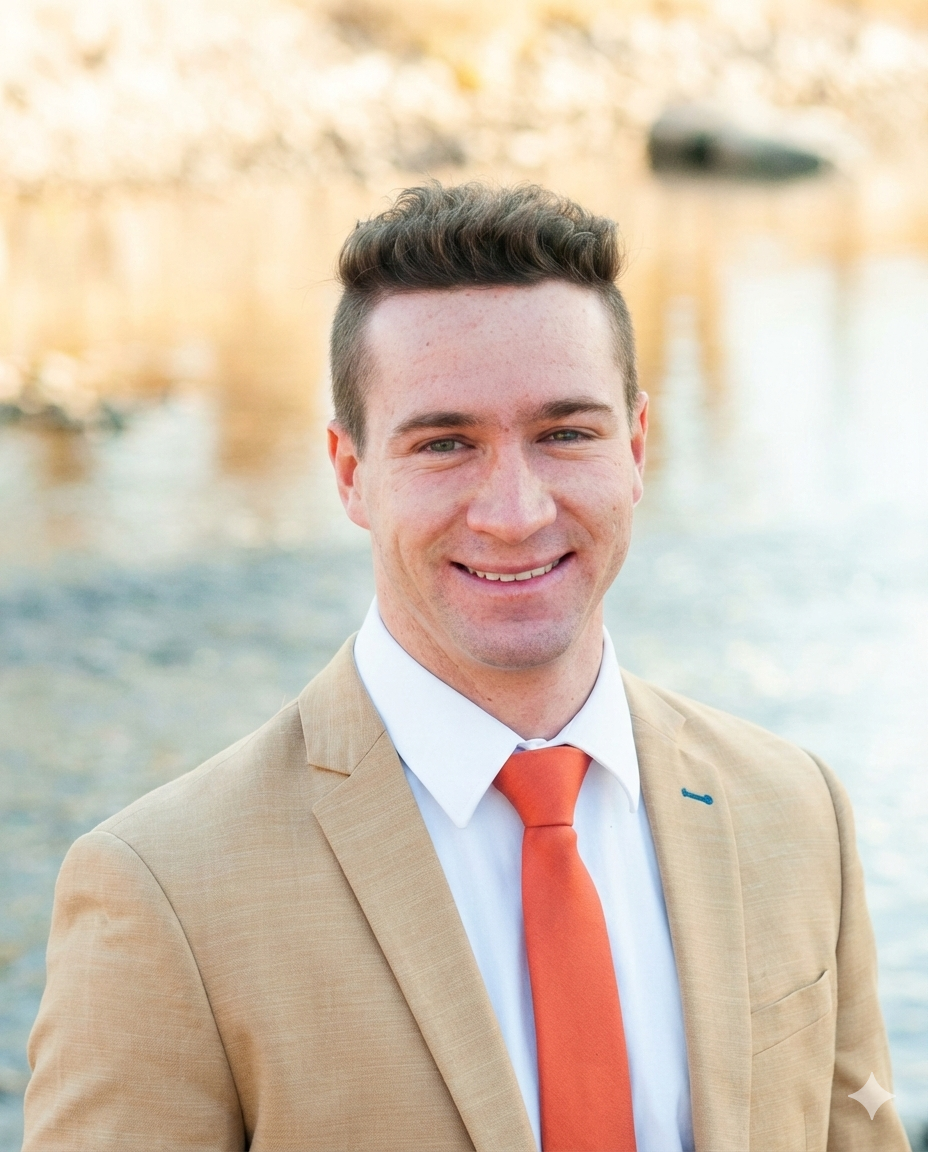}}]
{BRYTON PETERSEN}~received his B.Sc. degree in Computer Science from Brigham Young University-Idaho in 2022. He is currently a wireless security researcher at Idaho National Laboratory (INL), a U.S. Department of Energy national laboratory focused on energy innovation, security, and advanced technology. At INL, his efforts span machine learning and wireless systems research related to network security and traffic analysis. His contributions include co-development of publicly available datasets and machine-learning methods for 5G network traffic analysis and attack detection, advancing capabilities for secure communications in next-generation networks.
\end{IEEEbiography}

\begin{IEEEbiography}[{\includegraphics[width=1in,height=1.25in,clip,keepaspectratio]{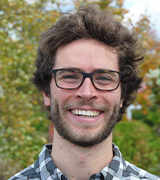}}]
{ANTOINE LESAGE-LANDRY}~(Senior~Member,~IEEE) is an Associate Professor in the Department of Electrical Engineering at Polytechnique Montréal, QC, Canada. He received the B.Eng. degree in Engineering Physics from Polytechnique Montréal, QC, Canada, in 2015, and the Ph.D. degree in Electrical Engineering from the University of Toronto, ON, Canada, in 2019. From 2019 to 2020, he was a Postdoctoral Scholar in the Energy \& Resources Group at the University of California, Berkeley, CA, USA. His research interests include optimization and machine learning, and their application to renewable power systems and wireless communications.
\end{IEEEbiography}

\begin{IEEEbiography}[{\includegraphics[width=1in,height=1.25in,clip,keepaspectratio]{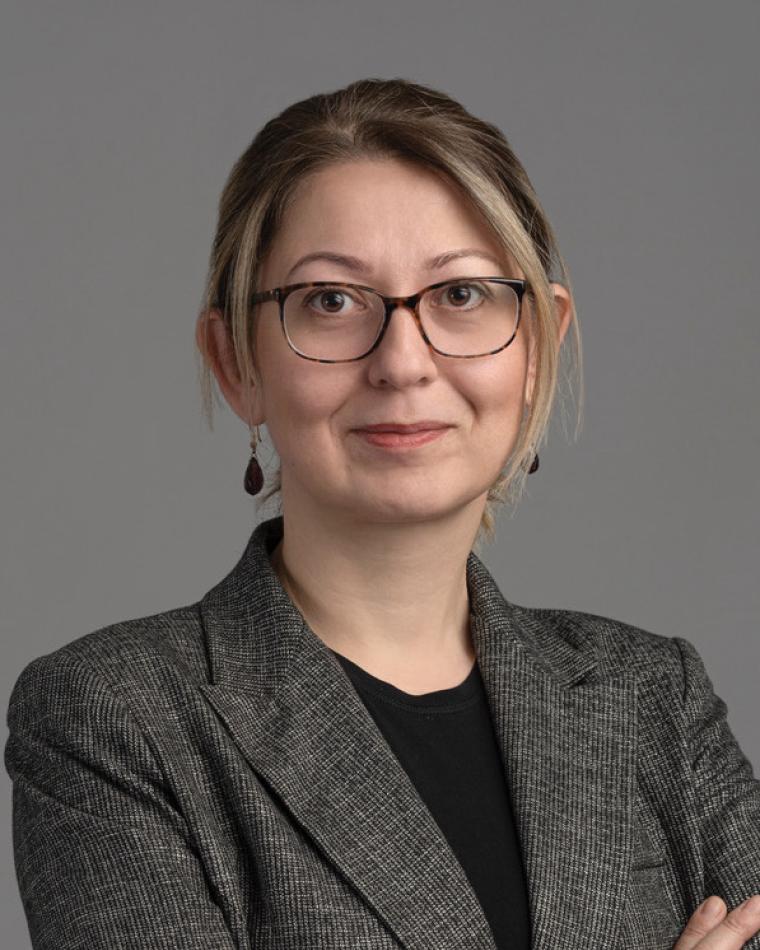}}]
{GÜNEŞ KARABULUT KURT}~(Senior~Member,~IEEE) is a Canada Research Chair (Tier 1) in New Frontiers in Space Communications and Full Professor at Polytechnique Montréal, Montréal, QC, Canada. She is also an adjunct research professor at Carleton University. Gunes received the B.S. degree with high honors in electronics and electrical engineering from Bogazici University, Istanbul, Turkiye, in 2000 and the M.A.Sc. and the Ph.D. degrees in electrical engineering from the University of Ottawa, ON, Canada, in 2002 and 2006, respectively. She worked in different technology companies in Canada and Turkiye between 2005 and 2010. From 2010 to 2021, she was a professor at Istanbul Technical University. Gunes is a Marie Curie Fellow and has received the Turkish Academy of Sciences Outstanding Young Scientist (TÜBA-GEBIP) Award in 2019. She is serving as the secretary of the IEEE Satellite and Space Communications Technical Committee,  the chair of the IEEE special interest group entitled “Satellite Mega-constellations: Communications and Networking,” and also as an editor in 6 different IEEE journals. She is a member of the IEEE WCNC Steering Board and a Distinguished Lecturer of the Vehicular Technology Society Class of 2022. Her research interests include multi-functional space networks, space security, and wireless testbeds.
\end{IEEEbiography}

\begin{IEEEbiography}[{\includegraphics[width=1in,height=1.25in,clip,keepaspectratio]{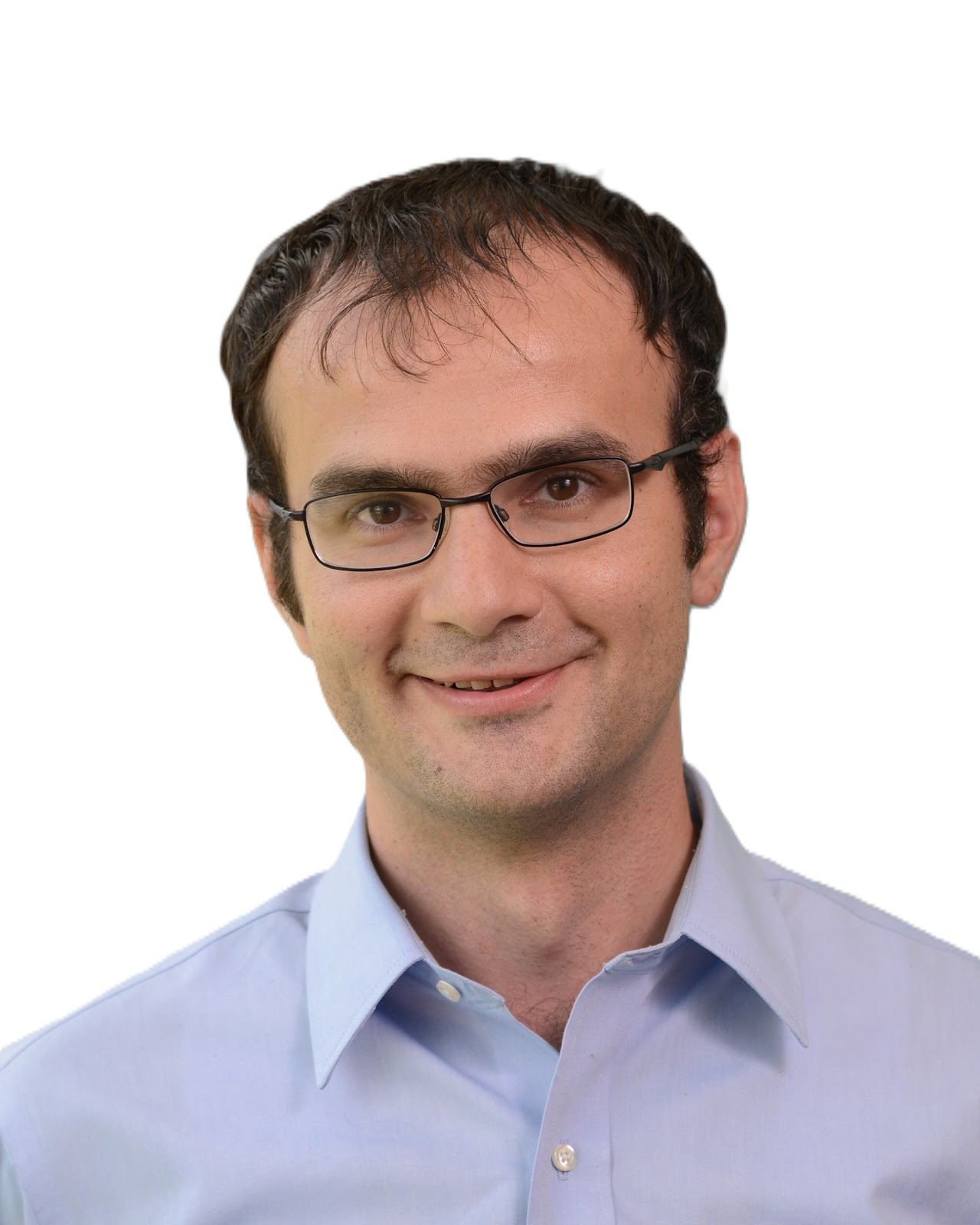}}]
{\.{I}SMA\.{I}L G{\"{U}}VEN{\c{C}}}~(Fellow,~IEEE) is a Professor at the Department of Electrical and Computer Engineering at NC State University. His recent research interests include 5G/6G wireless networks, UAV communications, millimeter/terahertz communications, and heterogeneous networks. He has published more than 300 conference/journal papers and book chapters, several standardization contributions, four books, and over 30 U.S. patents. Dr. Guvenc is the PI and the director for the NSF AERPAW project and a site director for the NSF BWAC I/UCRC center. He is an IEEE Fellow, a senior member of the National Academy of Inventors, and a recipient of several awards, including NC State University Alcoa Distinguished Engineering Research Award (2023), Faculty Scholar Award (2021), R. Ray Bennett Faculty Fellow Award (2019), FIU COE Faculty Research Award (2016), NSF CAREER Award (2015), Ralph E. Powe Junior Faculty Award (2014), and USF Outstanding Dissertation Award (2006).
\end{IEEEbiography}


\end{document}